\documentclass[journal]{IEEEtran}
\usepackage{stfloats}
\usepackage{algorithm}  
\usepackage{algpseudocode}  
\usepackage{amsmath}  

\usepackage{graphicx}
\usepackage[]{algpseudocode}
\usepackage{algorithmicx,algorithm}
\usepackage{amsmath}
\usepackage{titlesec}
\usepackage{multirow}  
\usepackage{amsthm,amsmath,amssymb}
\usepackage{graphicx}
\usepackage{float}
\usepackage{subfigure}
\usepackage{mathrsfs}
\usepackage{url}
\usepackage{array}
\newcolumntype{C}[1]{>{\centering}p{#1}}
\usepackage{makecell}
\usepackage{diagbox}
\usepackage{amsmath}
\newtheorem{remark}{Remark}
\usepackage{gensymb}
\usepackage{cite}
\usepackage{color}

\ifCLASSOPTIONcompsoc
 \usepackage[caption=false,font=normalsize,labelfont=sf,textfont=sf]{subfig}
\else
\usepackage[caption=false,font=footnotesize]{subfig}
\fi

\UseRawInputEncoding
\begin{document}
\vspace{-15mm}
\title{Near Field Computational Imaging with RIS Generated Virtual Masks}

\author{Yuhua Jiang, Feifei Gao, Yimin Liu, Shi Jin, and Tiejun Cui


\thanks{
Manuscript received 14-Apr-2023; revised 09-Sep-2023 and 08-Feb-2024;
accepted 05-Mar-2024. 
This work was supported by the 
National Natural Science Foundation of China under Grants 623B1013, 62325107, 62341107, and 62261160650. (Corresponding author: Feifei Gao.)

Y. Jiang, and F. Gao are with Institute for Artificial Intelligence, Tsinghua University (THUAI), 
State Key Lab of Intelligent Technologies and Systems, Tsinghua University, 
Beijing National Research Center for Information Science and Technology (BNRist), Beijing, P.R. China (email: jiangyh20@mails.tsinghua.edu.cn, feifeigao@ieee.org).

Y. Liu is with the Intelligent Sensing Laboratory,
Department of Electronic Engineering, Tsinghua University, Beijing 100084,
China (e-mail: yiminliu@tsinghua.edu.cn).

S. Jin is with the National Mobile Communications Research
Laboratory, Southeast University, Nanjing 210096, China (e-mail: jinshi@seu.edu.cn).

T. Cui is with the State Key Laboratory of Millimeter Waves, Southeast University, Nanjing 210096, China (e-mail: tjcui@seu.edu.cn).
}
}

\maketitle
\vspace{-15mm}
\begin{abstract}

Near field computational imaging has been recognized as a promising technique for non-destructive and highly accurate detection of the target.
Meanwhile, reconfigurable intelligent surface (RIS) can flexibly control the scattered electromagnetic (EM) fields for sensing the target and can thus help computational imaging in the near field.
\textcolor{black}{
In this paper, we propose a near-field imaging scheme based on holograghic aperture RIS for both 2D and 3D targets.
}
Specifically, we first establish an end-to-end EM propagation model from the perspective of Maxwell equations.
To mitigate the inherent ill conditioning of the inverse problem in the imaging system, we design the EM field patterns as masks that help translate the inverse problem into a forward problem. 
Next, we utilize RIS to generate different virtual EM masks on the target and calculate the cross-correlation between the mask patterns and the electric field strength at the receiver.
We then provide a RIS design scheme for virtual EM masks by employing a regularization technique.
The cross-range resolution of the proposed method is analyzed based on the spatial spectrum of the generated masks.
\textcolor{black}{
Simulation results demonstrate that the proposed method can produce satisfactory imaging results for both 2D and 3D targets. }%
Moreover, the imaging quality can be improved by generating more virtual EM masks, by increasing the signal-to-noise ratio (SNR) at the receiver, or by placing the target closer to the RIS.

\end{abstract}


\IEEEpeerreviewmaketitle

\section{Introduction} 






\textcolor{black}{
Electromagnetic (EM) imaging technology is non-contact, non-invasive, and could provide high-resolution images of the object \cite{Sheen:3DmmWaveImaging,9827908,8668512,9743695,4148072}.  
EM imaging can be used for inspection and visualization of the human body, which meets the increasing demands from social security, rescue activity, and medical applications. }%
Conventional EM imaging techniques include optimization-based imaging \cite{rubaek2007nonlinear}, confocal radar-based imaging \cite{li2005overview}, and EM tomography \cite{sadeghi2019dort}.
In optimization-based imaging, the exact nonlinear diffraction relation between the measured scattered field and the complex permittivity is found iteratively \cite{ali20103d}.
In confocal microwave imaging, a wide-band pulse is transmitted to detect strong scatterers inside the inspected region \cite{li2005overview}.
In EM tomography, the transmitted signals in multiple planes intersect the imaged object and are used to reconstruct 2D slice images in different planes \cite{ren2016uniform}. 
\textcolor{black}{
However, conventional EM imaging typically demands changeable EM sources such as scanning antennas and dynamic metasurface antennas \cite{scan,molaei2022development}. 
When the EM source is unchangeable, conventional EM imaging methods may not be applicable.
In this case, reconfigurable intelligent surface (RIS) has recently been used to reflect EM waves from the source and 
generate multiple probing signals in imaging systems \cite{RIS_image}.
}
The use of RIS for imaging has several advantages.
Firstly, RIS is a flexible and cost-effective solution that can be easily integrated into existing imaging systems \cite{9384499,9149203,8746155,9145334}.
Secondly, RIS can operate over a wide range of frequencies, e.g., from GHz waves to THz waves, making it suitable for a variety of applications \cite{5,6,7,8,RIS,9343768}. 
\textcolor{black}{
Thirdly, by precisely controlling the wavefront of EM waves, RIS can potentially enhance the signal-to-noise ratio (SNR) and the resolution of the imaging system \cite{RIS_image,near_imaging,9,10,mypaper}.
}


\textcolor{black}{
However, most existing literature about near-field imaging applies discrete aperture RIS, which can not achieve high spatial resolution and energy efficiency.
Holographic aperture RIS, in particular, could alter the reflection coefficients continuously and could generate complex wavefronts with a high degree of control and flexibility \cite{h1,h2,antenna}. As a result, the holographic aperture RIS can potentially provide even higher resolution and accuracy in near-field imaging \cite{h3}, \cite{h4}.
}
Despite these promising advantages, the use of holographic RIS for near-field imaging is still a relatively new area and faces many challenges, including optimizing the design and configuration of the holographic RIS to achieve the desired imaging performance, as well as developing effective signal processing algorithms to extract useful information from the received signals.
A more intractable challenge is the inherent ill conditioning of the inverse problem in the imaging system, which brings numerical errors and large computation overhead.
Therefore, while there is much excitement surrounding the potential applications of holographic RIS in near-field imaging, more investigation is needed to fully understand its capabilities and restrictions.

\textcolor{black}{
In this paper, we develop a novel imaging scheme that utilizes RIS to build virtual EM masks on the target. 
The method is applicable to both 2D and 3D objects.
}
Specifically, we first establish an end-to-end EM propagation model from the perspective of Maxwell equations.
To mitigate the inherent ill conditioning of the inverse problem in the imaging system \cite{inverse}, we design the EM field patterns as masks that help translate the inverse problem into a forward problem.
Next, we utilize RIS to generate different virtual EM masks on the target and calculate the cross-correlation between the mask patterns and the electric field strength at the receiver.
We then provide a RIS design scheme for virtual EM masks by employing a regularization technique.
The cross-range resolution of the proposed method is analyzed based on the spatial spectrum of the generated masks.
\textcolor{black}{
The proposed method translates the original inverse problem into a forward problem and has the advantage of simplifying the imaging process and reducing the computation complexity.
}
Simulation results demonstrate that the proposed method can achieve high-quality imaging.
Moreover, the imaging quality can be improved by generating more virtual EM masks, by increasing the SNR at the receiver, or by placing the target closer to the RIS.


The rest of this paper is organized as follows. Section~\uppercase\expandafter{\romannumeral2}
presents the system model and formulates the imaging problem.
Section~\uppercase\expandafter{\romannumeral3} derives the end-to-end electromagnetic propagation formula.
Section~\uppercase\expandafter{\romannumeral4} elaborates the strategy of reconstructing the image with multiple measurements.
Section~\uppercase\expandafter{\romannumeral5} describes the virtual EM masks design criterion and the way to generate those masks by the RIS.
Section~\uppercase\expandafter{\romannumeral6} provides the numerical simulation results, and 
Section~\uppercase\expandafter{\romannumeral7} draws the conclusion.

Notations: Boldface denotes vector or matrix;  $j$ corresponds to the imaginary unit; $(\cdot)^H$, $(\cdot)^T$, and $(\cdot)^*$ represent the Hermitian, transpose, and conjugate, respectively; 
 $\nabla$ is the nabla operator; $\nabla^2$ is the Laplacian operator;
 $\delta(\cdot)$  is the Dirac delta function; 
 $\delta_{nm}$ is the Kronecker delta function; 
 $\left| a \right|$ and $\angle a$ denote the length and angle of the complex number $a$, respectively;
$\left|\mathbf{a}\right|$ denotes the vector composed of lengths of each element in complex vector $\mathbf{a}$; $\left\Vert\mathbf{a}\right\Vert$ denotes 2-norm of the vector $\mathbf{a}$; $\left\Vert\mathbf{A}\right\Vert_F$ denotes Frobenius-norm of the matrix $\mathbf{A}$; $\left\langle \cdot \right\rangle_I$ denotes the ensemble average over a series of $I$ measurements. 

\section{System Model and Problem Formulation}
\subsection{System Model}
\begin{figure}[t]
  \centering
  \centerline{\includegraphics[width=7.7cm]{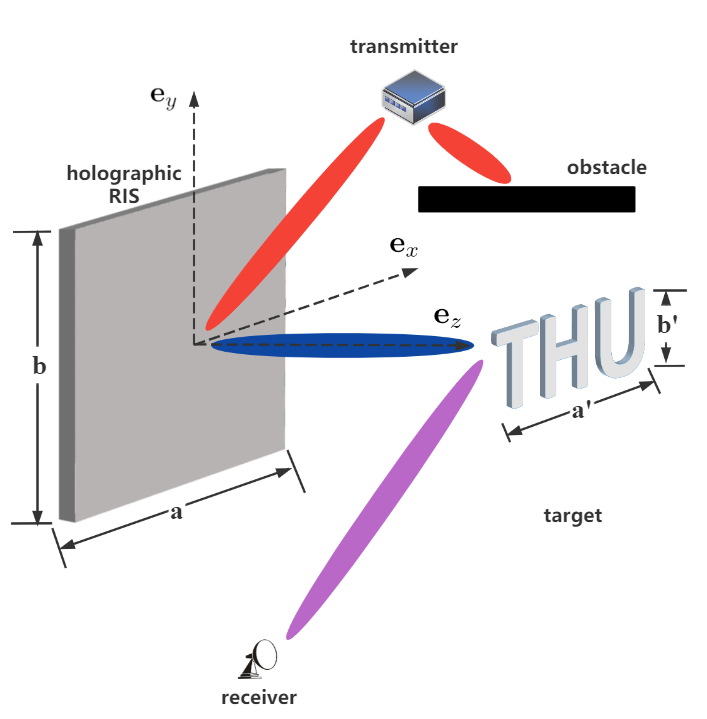}}
  \caption{The imaging system model is composed of a transmitter, a RIS, a target, and a receiver. 
  }
  \label{system_model}
\end{figure}
As shown in Fig.~1, we consider a holographic RIS-aided computational imaging system, which consists of a
single-antenna transmitter, a holographic RIS, a 2D target, and a single-antenna
receiver.
\textcolor{black}{
Suppose the transmitter is located in the far field of the RIS, such that the incident waves on the RIS becomes planar waves. 
Assume the receiver is located in the far field of the target and detects the electrical intensity of the EM waves scattered by the target. 
}
Since the detailed patterns of the target can not be recognized in the far field of the RIS due to the resolution limit, the target should be placed in the near field. 
Meanwhile, we assume that there is an obstacle between the transmitter and the target, which is thick and is composed of dielectric, such that the EM waves transmitted through the obstacle can be neglected.

Specifically, we consider the RIS as a rectangular plane $S_1$ with length $a$ in $x$-axis and length $b$ in $y$-axis, whose center is located at the origin. Suppose the wave number is $k=2\pi/\lambda$ where $\lambda$ is the wavelength. 
As in \cite{9424177}, \cite{9110889}, the holographic RIS can realize a continuous reflection with coefficient function $\Gamma(x, y) = \tau(x,y) e^{j\beta(x,y)}$, where $\tau(x,y)$ is the amplitude coefficient and $\beta(x, y)$ is the phase shift at the point $(x, y, 0)$ on the RIS.
Denote the distance from the target to the RIS as $z'$ and denote $(x',y')$ as the horizontal coordinate when $z=z'$.
Moreover, the lengths of the target in $x$-axis and in $y$-axis are $a'$ and $b'$, respectively. 
Normally, the target is relatively small in size compared to RIS, i.e., $a'<a$ and $b'<b$.



The RIS is considered active, which has been introduced in prior literature about wireless communication systems \cite{active1}, \cite{active2}. 
Different from passive RIS that passively reflects signals without amplification, the key feature of active RIS is their ability to actively reflect signals with amplification, which can be realized by integrating
reflection-type amplifiers into their reflecting elements \cite{active3}, \cite{active4}.
Note that, in active RIS it is possible that $\tau(x,y)>1$ thanks to the integrated
reflection-type amplifier. Denote the total amplification rate at RIS as $P_I$, 
which is the ratio of the power of the incident wave on the RIS to the power of the reflected wave on the RIS. Thus, the overall power restriction of the active RIS can be expressed as
\begin{align}
\iint_{S_1} |\Gamma(x, y)|^2 dxdy \leq ab P_I.
\label{conti}
\end{align}

For 2D target reconstruction, we assume that the target is thin, i.e., its dimension
along the $z$ direction is much smaller than $a'$ and $b'$.
The obvious restriction here is that, the target must lie in a virtual plane more or less parallel to the RIS for proper reconstruction \cite{2d}.
Specifically, we place the target on the virtual plane $S_2$ parallel to $xy$ plane whose center is located at $(0,0,z')$. 
The scattering density of the target can be described by the function $T(x', y')$, which is in the form of a piecewise constant function.
If the target covers the point $(x', y', z')$, then $T(x', y')=1$; otherwise $T(x', y')=0$.
Since the target is in the near field of the RIS, then there is
\cite{soft}:
\begin{align}
z' < \frac{2 (a^2+b^2)}{\lambda}.
\label{rayleigh}
\end{align}
The upper bound of the near-field region, i.e., the right hand side of (\ref{rayleigh}) is called the Rayleigh distance for the RIS. 
\textcolor{black}{
Let $d_t$ denote the distance between the transmitter and the center of the RIS, and 
let $d_r$ denote the distance between the receiver and the center of the target.   
Then the far field restriction for the locations of the transmitter and the receiver are respectively given by \cite{soft}:
\begin{align}
d_t > \frac{2 (a^2+b^2)}{\lambda}, \quad
d_r > \frac{2 (a'^2+b'^2)}{\lambda}.
\end{align}
}


\subsection{Formulation for Imaging Problem}
The imaging problem is equivalent to the estimation of $T(x', y')$.
Since it is hard to capture $T(x', y')$ on the target all at once,
we should produce multiple measurements by altering the reflection coefficients on the RIS. 
In different measurements, the reflected waves cause various EM field patterns on the target surface to capture different target features.
Since the EM field patterns encrypt $T(x', y')$, we refer to these EM field patterns as virtual EM masks.
In order to encrypt $T(x', y')$ efficiently, the virtual EM masks should be elaborately designed.
The designs of the virtual EM masks and the corresponding RIS coefficients are conducted offline before the measurements.
In the process of the measurements, the receiver records the electrical intensity corresponding to different virtual EM masks.
After all measurements are completed, $T(x', y')$ can be estimated at the receiver on the basis of the correlation between 
virtual mask patterns and the received electrical intensity.





\section{Modeling the End-to-End Electromagnetic Propagation}
\subsection{Scattering on the RIS}
We assume the EM field configuration of the incident wave is the transverse magnetic $\mathbf{e}_x$, i.e., the E-field is parallel to $\mathbf{e}_x$ while the H-field lies in the plane spanned by $\mathbf{e}_y$ and $\mathbf{e}_z$. Since the transmitter is located in the far field, the EM waves incident on the RIS are considered as planar waves that have the electric and the magnetic field distributions:
\begin{align}
\mathbf{E}_{\text{in}}   &=E_{0} e^{-j k\left(\sin \left(\theta_{\text{in}}\right) y-\cos \left(\theta_{\text{in}}\right) z\right)} \boldsymbol{e}_{x}, \\
\mathbf{H}_{\text{in}} &=-\frac{E_{0}}{\eta}\left(\cos \left(\theta_{\text{in}}\right) \boldsymbol{e}_{y}+\sin \left(\theta_{\text{in}}\right) \boldsymbol{e}_{z}\right) e^{-j k\left(\sin \left(\theta_{\text{in}}\right) y-\cos \left(\theta_{\text{in}}\right) z\right)},
\end{align}
where $\eta=\sqrt{\frac{\mu}{\varepsilon}}$ is the characteristic impedance of the medium, while $\mu$ and $\varepsilon$ separately represent the magnetic permeability and the electric permittivity.

The outgoing EM waves scattered by the RIS are defined as $\mathbf{E}_{\text{out}}$ and  $\mathbf{H}_{\text{out}}$, respectively.
According to the EM reflection theorem \cite{textbook}, $\mathbf{E}_{\text{out}}$ and  $\mathbf{H}_{\text{out}}$ satisfy the relations
\begin{align}
\left.\mathbf{E}_{\text{out}}  \right|_{z=0}&=\left.\Gamma(x, y) \mathbf{E}_{\text{in}}  \right|_{z=0} \\
\mathbf{e}_{z} \times\left.\mathbf{H}_{\text{out}}\right|_{z=0}&=-\Gamma(x, y) \mathbf{e}_{z} \times\left.\mathbf{H}_{\text{in}}\right|_{z=0}.
\end{align}
Since the thickness of the RIS is always negligible compared to its length and width,
we suppose that the RIS can be replaced by an imaginary surface.
Denote the transmitted fields below the imaginary 
surface by $\mathbf{E}_{\text{tr}}$ and $\mathbf{H}_{\text{tr}}$, respectively. 
According to the physical equivalent theorem \cite{textbook}, $\mathbf{E}_{\text{in}}  $ and $\mathbf{H}_{\text{in}}$ above the imaginary surface can be removed.
Then, an equivalent electric current density $\mathbf{J}_{e}$ and a magnetic current density $\mathbf{M}_{e}$ should be imposed on the imaginary surface to satisfy the boundary conditions \cite{textbook}, which can be separately expressed as
\begin{align}
\mathbf{J}_{e} &=\mathbf{e}_{z} \times\left(\left.\mathbf{H}_{\text{out}}\right|_{z=0}-\left.\mathbf{H}_{\text{tr}}\right|_{z=0}\right), \\
\mathbf{M}_{e} &=-\mathbf{e}_{z} \times\left(\left.\mathbf{E}_{\text{out}}  \right|_{z=0}-\left.\mathbf{E}_{\text{tr}}\right|_{z=0}\right).
\end{align}

Conventionally, the RIS can be assumed as a perfect electrical conducting (PEC) surface \cite{6}, where $\mathbf{E}_{\text{tr}}$, $\mathbf{H}_{\text{tr}}$, and $\mathbf{M}_{e}$ become zero.
Hence, only $\mathbf{J}_{e}$ contributes to the scattered fields. 
Next, the physical equivalent theory is applied to remove the PEC surface and to derive an unbounded environment \cite{textbook}. 
The equivalent electric current density $\mathbf{J}_{f}$ is expressed as 
\begin{align}
\mathbf{J}_{f}&=2 \mathbf{J}_{e}=2 \Gamma(x,y) \mathbf{e}_{z} \times\left.\mathbf{H}_{\text{in}}\right|_{z=0} \nonumber \\
&=2\frac{E_0}{\eta} \cos \theta_{\text{in}} e^{-j k \sin \left(\theta_{\text{in}}\right) y} \Gamma(x,y) \mathbf{e}_{x}\overset{\Delta}{=}J_{x} \Gamma(x,y) \mathbf{e}_{x}. \hfill
\label{start}
\end{align}


We then calculate the scattered EM fields of the RIS by first computing the magnetic vector potential, denoted as $\textbf{A}$.
In order to derive the relationship between $\textbf{A}$ and $\mathbf{J}_{f}$ in a concise form, we apply Lorenz Gauge \cite{textbook} throughout this paper. Then the following expression holds true for the magnetic vector potential
$\mathbf{A}$ at a near-field observation point on the target $(x^{\prime} , y^{\prime} , z^{\prime})$ \cite{textbook}:
\begin{align}
\mathbf{A}=\frac{\mu}{4 \pi} \iint_{S_1} \mathbf{J}_{f} \frac{e^{-j k R}}{R} d x d y,
\label{A1}
\end{align}
where $R=\sqrt{\left(x^{\prime}-x\right)^{2}+\left(y^{\prime}-y\right)^{2}+\left(z^{\prime}\right)^{2}}$ is the distance
from point $(x, y, 0)$ to $(x', y', z')$. Then, $\mathbf{H}_{\text{out}}$ can be derived as 
\begin{align}
\mathbf{H}_{\text{out}}&= \frac{1}{\mu}\nabla \times \mathbf{A}
=\frac{1}{4 \pi} \iint_{S_1} \nabla \times \left(\frac{e^{-j k R}}{R} \mathbf{J}_{f} \right)  d x d y.
\label{12}
\end{align}


In order to simplify the integrand in (\ref{12}), we utilize the vector identity and obtain
\begin{align}
\nabla \times  \left(\frac{e^{-j k R}}{R} \mathbf{J}_{f} \right)&=\nabla \left(\frac{e^{-j k R}}{R}\right) \times J_x \Gamma(x,y) \mathbf{e}_x\nonumber \\
&+\frac{e^{-j k R}}{R}\left(\nabla \times J_x \Gamma(x,y) \mathbf{e}_x\right).
\end{align}
Since the nabla operators take the partial derivative with respect to $x'$, $y'$, and $z'$, $\nabla \times J_x \Gamma(x,y) \mathbf{e}_x$ equals $0$.
Thus, we can compute the three rectangular components of $\mathbf{H}_{\text{out}}$ as
\begin{align}
H_{\text{out}}^x & =0 ,\\
H_{\text{out}}^y & =-\frac{1}{4 \pi} \iint_{\mathcal{S}_1} \frac{1+j k R}{ R^3} z' J_x \Gamma(x,y) e^{-j k R} d x d y, \\
H_{\text{out}}^z & =-\frac{1}{4 \pi} \iint_{\mathcal{S}_1} \frac{1+j k R}{ R^3}\left(y-y'\right) J_x \Gamma(x,y) e^{-j k R} d x d y.
\label{H3}
\end{align}

\textcolor{black}{Note that, only when the target is placed in the near field of RIS would $\mathbf{H}_{\text{out}}$ varies significantly on different points of the target.
Thus, $\mathbf{H}_{\text{out}}$ can be used as masks to detect the target and then the imaging of the target can be achieved.}
In order to capture possibly more information of the target, we need to elaborately design the EM field distribution as virtual masks, in which we need to find out a proper distribution of $\Gamma(x,y)$ that can generate the desired EM fields.
\textcolor{black}{
However, calculating the distribution of $\Gamma(x,y)$ with a given $H_{\text{out}}^z$ or $H_{\text{out}}^y$ is an inherently ill-conditioned inverse problem, where small perturbations in the measured field data cause large perturbations in the recovered distributions of $\Gamma(x,y)$ \cite{inverse}. 
}
In general, we utilize the \textit{method of moments} (MoM) \cite{textbook} to discretize the continuous function of $\mathbf{H}_{\text{out}}$ over both $S_1$ and $S_2$. 
Discretization has a regularizing effect in the sense that the discrete formulation admits a bounded inverse though it is highly unstable.
Therefore, discrete regularization is a necessary procedure in order to recover a
physically meaningful solution.

The discretization process is conducted by sampling on $S_1$ and on $S_2$, 
where the numbers of sampled points are denoted by $N$ and $M$, respectively. Define $\Delta_x$ and $\Delta_y$ as the intervals of the sampled points on $S_1$  in $x$ and $y$ directions, respectively.
Denote  $\mathbf{r}_n=(x_n,y_n ,0 )^T$ as the $n$th sampled point on $S_1$ and 
denote $\mathbf{r}_m'=(x_m',y_m' ,z' )^T$ as the $m$th field point on $S_2$.
Since only the tangential components of $\textbf{H}_{\text{out}}$ will induce electric current on the target and since $H_{\text{out}}^x =0$, we only focus on the discretization of $H_{\text{out}}^y$.
The formula in (\ref{H3}) can be simplified to
\begin{equation}
H_{\text{out}}^y\left(\mathbf{r}_m'\right)=\sum_{n=1}^N Z\left(\mathbf{r}_m', \mathbf{r}_n\right) \Gamma\left(\mathbf{r}_n\right), \quad m=1, \cdots, M ,
\label{moments}
\end{equation}
where $Z\left(\mathbf{r}_m', \mathbf{r}_n\right)=-  \left. \frac{1+j k R}{4 \pi R^3} \Delta_x \Delta_y z' J_x e^{-j k R}\right|_{\mathbf{r}_m', \mathbf{r}_n}$ represents the discretized integral kernel. 
Denote $\textbf{y}=[H_{\text{out}}^y\left(\mathbf{r}_1'\right),\cdots,H_{\text{out}}^y\left(\mathbf{r}_M'\right)]^T$, $\textbf{p}=[\Gamma\left(\mathbf{r}_1\right),\cdots,\Gamma\left(\mathbf{r}_N\right)]^T$, and let $\textbf{Z} \in \mathbb{C}^{M\times N}$ collect all $Z\left(\mathbf{r}_m', \mathbf{r}_n\right)$.
Then, (\ref{moments}) can be further written as
\begin{equation}
\textbf{y}= \textbf{Z} \textbf{p} .
\label{linear}
\end{equation}
\textcolor{black}{
Note that, $\textbf{y}$ needs to be elaborately designed for the imaging problem.
}
With the discrete form, the power restriction in (\ref{conti}) should be expressed as
\begin{equation}
\| \textbf{p} \|^2 \le N P_I.
\label{disc}
\end{equation}

\subsection{Scattering on a 2D Homogeneous Target}
\textcolor{black}{
To perform the near field computational imaging, the concrete expression of the EM waves scattered from the target must be determined.
Hence, in this subsection, we will analyze the scattering process on a 2D homogeneous target, taking the scattered waves from the RIS as the incident waves.
} %

We first replace the role of the target with equivalent electric current density $\textbf{J}_f'$. 
Denote the reflected magnetic field of $\mathbf{H}_{\text{out}}$ as $\mathbf{H}_{\text{r}}$.
Then, the physical equivalent current density on the surface of the target is
\begin{align}
\mathbf{J}_{f}'= - \mathbf{e}_{z} \times \left.\mathbf{H}_{\text{total}}\right|_{z=z'}
= - \mathbf{e}_{z} \times (\left.\mathbf{H}_{\text{out}}\right|_{z=z'}+\left.\mathbf{H}_{\text{r}}\right|_{z=z'}).
\end{align}
Here $\mathbf{H}_{\text{r}}$ can be represented as $\mathbf{H}_{\text{r}}=-\Gamma' \mathbf{H}_{\text{out}}$, where $\Gamma'$ is the reflection coefficient of the surface. 
Then $\textbf{J}_f'$ can be written as
\begin{align}
\mathbf{J}_{f}'&= 
-(1-\Gamma') \mathbf{e}_{z} \times \left.\mathbf{H}_{\text{out}}\right|_{z=z'}=(1-\Gamma')\left.H_{\text{out}}^y\right|_{z=z'} \mathbf{e}_{x} \nonumber \\
&\overset{\Delta}{=}J_{x}' (x',y') \mathbf{e}_{x}.
\label{twice}
\end{align}
\textcolor{black}{
From the fact $H_{\text{out}}^x=0$ and $H_{\text{out}}^y\neq0$, we assert that $\mathbf{J}_{f}'$ only has the component in $\mathbf{e}_{x}$ direction.
}
The reflection coefficient $\Gamma'$ is determined by the surface impedance and can be assumed as a constant across the surface of the target.
Specifically, $\Gamma'$ is equal to $-1$ for a PEC surface.

If we want to determine the scattered fields of the target, then the vector potential
$\textbf{A}_r$ at the receiver $\textbf{r}_r=(x_r,y_r,z_r)$ should be written as \cite{textbook}
\begin{align}
\mathbf{A}_r=\frac{\mu}{4 \pi} \iint_{S_2} T(x',y') \mathbf{J}_{f}' \frac{e^{-j k R'}}{R'} d x' d y',
\label{a1}
\end{align}
where $R'=\sqrt{\left(x^{\prime}-x_r\right)^{2}+\left(y^{\prime}-y_r\right)^{2}+\left(z^{\prime}-z_r\right)^{2}}$ is the distance
from point $(x', y', z')$ to $(x_r,y_r,z_r)$.
\textcolor{black}{
Note that $\mathbf{A}_r$ is intrinsically influenced by the target scattering density $T(x',y')$ because the equivalent current can only exist on the area that is covered by the target.
}
Then, $\mathbf{E}_{r}$ can be derived as 
\begin{align}
\mathbf{E}_{r}&=\frac{1}{j k \sqrt{\mu \varepsilon}} \nabla \times(\nabla \times \mathbf{A}_r)=\frac{1}{j k \sqrt{\mu \varepsilon}} \left(\nabla(\nabla \cdot \mathbf{A}_r)-\nabla^2\mathbf{A}_r\right)\nonumber \\
&=\frac{1}{j k \sqrt{\mu \varepsilon}} \left(\nabla(\nabla \cdot \mathbf{A}_r)+k^2\mathbf{A}_r\right).  
\label{a2}
\end{align}
The last equality in (\ref{a2}) holds because with the absence of the electric current source, $\mathbf{A}_r$ satisfies the vector Helmholtz equation $\nabla^2\mathbf{A}_r+k^2\mathbf{A}_r=\textbf{0}$. 
For the last equality in (\ref{a2}), $\nabla(\nabla \cdot \mathbf{A}_r)$
only contributes variations of the order $\mathcal{O}(\frac{1}{R'^2}),\mathcal{O}(\frac{1}{R'^3}),\mathcal{O}(\frac{1}{R'^4})$, etc,
while $k^2\mathbf{A}_r$ contains variation of the order $\mathcal{O}(\frac{1}{R'})$ \cite{textbook}.
Considering that the receiver is located in the far field of the target, the dominant variation is of the order $\frac{1}{R'}$ and is contained in the $k^2\mathbf{A}_r$ term. Thus, when the receiver is located in the far field of the target, (\ref{a2}) reduces to
\begin{align}
&\mathbf{E}_{r}\!\approx\! \frac{k}{j \sqrt{\mu \varepsilon}}\mathbf{A}_r\nonumber\!= \!\frac{k \sqrt{\mu}}{4 \pi j \sqrt{\varepsilon}} \!\iint_{S_2} \!\!\!\frac{e^{-j k R'}}{R'} T(x',y') J_{x}' (x',y') \mathbf{e}_{x} d x' d y' \nonumber \\
&\overset{\Delta}{=} \iint_{S_2} K(x',y') T(x',y') J_{x}' (x',y') \mathbf{e}_{x} d x'\! d y',
\label{reduce}
\end{align}
where $K(x',y') \overset{\Delta}{=} \frac{k \sqrt{\mu}}{4 \pi j \sqrt{\varepsilon}} \frac{e^{-j k R'}}{R'}$ is the point spread function (PSF).
Substituting $J_x'=2\left.H_{\text{out}}^y\right|_{z=z'}$ into (\ref{reduce}), we can further compute $\mathbf{E}_{r}$ with the knowledge of $\textbf{H}_{\text{out}}$, which is determined by $\Gamma(x,y)$.
Since $\mathbf{E}_{r}$ only consists of the $x$-axis component, the scalar electric intensity at the receiver can be calculated as
\begin{equation}
|E_r^x|=\left|\iint_{S_2} K(x',y') T(x',y') J_{x}' (x',y') d x' d y' \right| .
\end{equation}
When the incident wave on the RIS is fixed, $|E_r^x|$ is a function of $\Gamma(x,y)$ and $T(x', y')$.
Thus, the end-to-end electromagnetic propagation mechanism can be determined.



\textcolor{black}{
\subsection{Scattering on a 3D Inhomogeneous Target}
In this subsection, we will analyze the scattering process on a 3D inhomogeneous dielectric target.
With the equality $\mathbf{E}_{out}=-j \omega \mathbf{A}-j \frac{1}{\omega \mu \varepsilon} \nabla(\nabla \cdot \mathbf{A})$ and (\ref{A1}), we can derive the three rectangular components of $\mathbf{E}_{\text{out}}$ as (\ref{e1})-(\ref{e3}) at the top of this page.
\begin{figure*}[t]
\textcolor{black}{
\begin{align}
& E_{out}^x=- \frac{j \eta}{4 \pi k} \iint_{\mathcal{S}_1}\left[\frac{-1-jkR+k^2R^2}{R^3}+\frac{3+3 j kR-k^2R^2}{R^5}\left(x'-x\right)^2\right] J_x \Gamma(x,y) e^{-j kR} d x d y,  \label{e1}\\ 
& E_{out}^y=-\frac{j \eta}{4 \pi k} \iint_{\mathcal{S}_1} \frac{3+3 j kR-k^2R^2}{R^5}\left(y'-y\right)\left(x'-x\right) J_x \Gamma(x,y) e^{-j kR} d x d y, \label{e2}\\
& E_{out}^z=-\frac{j \eta}{4 \pi k} \iint_{\mathcal{S}_1} \frac{3+3 j kR-k^2R^2}{R^5} z'\left(x'-x\right) J_x \Gamma(x,y) e^{-j kR} d x d y . \label{e3}%
\end{align}%
\hrulefill
}%
\end{figure*} %
}

\textcolor{black}{
Assume prior knowledge has determined that the 3D target is contained within a detection region $D$. For 3D target imaging, it is equivalent to reconstructing the difference between the complex relative permittivity and the complex relative permittivity of air at each point $\mathbf{r}^{\prime}$ in region $D$, denoted as $\chi(\mathbf{r}^{\prime})$. Since the complex relative permittivity of air is approximately equal to $1$, we can formulate the contrast function as $\chi(\mathbf{r}^{\prime})=\epsilon_r(\mathbf{r}^{\prime})+j \sigma(\mathbf{r}^{\prime}) / (\epsilon_0 \omega)-1$, where $\epsilon_r(\mathbf{r}^{\prime})$ is the real relative permittivity at point $\mathbf{r}^{\prime}$, $\sigma(\mathbf{r}^{\prime})$ is the conductivity at point $\mathbf{r}^{\prime}$, $\epsilon_0$ is the vacuum permittivity,
and $\omega=2\pi f$ is the angular frequency of EM waves. 
The distribution of $\chi(\mathbf{r}^{\prime})$
is the target that needs to be recovered in imaging problems. 
For a 3D inhomogeneous dielectric target, the scattered field at the receiver can be expressed by the following Lippmann-Schwinger equation \cite{lipp}, \cite{lipp2}
\begin{equation}
\mathbf{E}_r\left(\mathbf{r}_r, \mathbf{r}^{\prime}\right)=k^2 \iiint_D \overline{\mathbf{G}}\left(\mathbf{r}_r, \mathbf{r}^{\prime}\right) \chi\left(\mathbf{r}^{\prime}\right) \mathbf{E}_{t o t}\left(\mathbf{r}^{\prime}\right) d \mathbf{r}^{\prime},
\label{gre}%
\end{equation}
where $\overline{\mathbf{G}}\left(\mathbf{r}_r, \mathbf{r}^{\prime}\right)$ is the tensor electric field Green function formulated as 
\begin{align}
&\overline{\mathbf{G}}\left(\mathbf{r}_r,\mathbf{r}^{\prime}\right)=\left(\mathbf{I}+\frac{\nabla \nabla}{k^2}\right) g\left(\mathbf{r}_r, \mathbf{r}^{\prime}\right)\nonumber\\
&=\!\!\left[\left(\frac{3}{k^2 R'^2}+\frac{3 j}{k R'}-1\!\!\right) \hat{\textbf{R}} \hat{\textbf{R}}^\top\!\!\!\!-\!\!\left(\frac{1}{k^2 R'^2}+\frac{j}{k R'}-1\!\!\right)\!\! \mathbf{I}\right] \!\!g\left(\mathbf{r}, \mathbf{r}^{\prime}\right). \label{RR} %
\end{align}
In (\ref{RR}), $\hat{\textbf{R}}$ is the unit vector from $\mathbf{r}^{\prime}$ to $\mathbf{r}_r$, and the scalar Green function is defined as $g\left(\mathbf{r}_r, \mathbf{r}^{\prime}\right) = \frac{e^{j k R'}}{4 \pi R'}$.
The tensor $\overline{\mathbf{G}}\left(\mathbf{r}_r, \mathbf{r}^{\prime}\right)$ could be rewritten in a matrix form, i.e.,
\begin{align}
\overline{\mathbf{G}}\left(\mathbf{r}_r, \mathbf{r}^{\prime}\right)
= 
\left[\begin{array}{lll}
G_{x x}\left(\mathbf{r}_r, \mathbf{r}^{\prime}\right) & G_{x y}\left(\mathbf{r}_r, \mathbf{r}^{\prime}\right) & G_{x z}\left(\mathbf{r}_r, \mathbf{r}^{\prime}\right) \\ 
G_{y x}\left(\mathbf{r}_r, \mathbf{r}^{\prime}\right) & G_{y y}\left(\mathbf{r}_r, \mathbf{r}^{\prime}\right) & G_{y z}\left(\mathbf{r}_r, \mathbf{r}^{\prime}\right) \\ 
G_{z x}\left(\mathbf{r}_r, \mathbf{r}^{\prime}\right) & G_{z y}\left(\mathbf{r}_r, \mathbf{r}^{\prime}\right) & G_{z z}\left(\mathbf{r}_r,\mathbf{r}^{\prime}\right) 
\end{array}\right].
\end{align}
}



\textcolor{black}{
Considering the Born approximation (BA) widely used in imaging problems \cite{born}, we assume $\mathbf{E}_{tot}=\mathbf{E}_{out}$.
Suppose the receiver can only measure the scalar electric field component in the $x$ direction, 
which can be formulated according to (\ref{gre}) and based on BA as  
\begin{align}
& E_r^x = k^2 \iiint_D  \chi\left(\mathbf{r}^{\prime}\right) \left[{G}_{xx}\left(\mathbf{r}_r, \mathbf{r}^{\prime}\right) E^x_{out}\left(\mathbf{r}^{\prime}\right) \right. 
 \nonumber\\
& \left.   
+ G_{xy}\left(\mathbf{r}_r, \mathbf{r}^{\prime}\right) E^y_{out}\left(\mathbf{r}^{\prime}\right)
\!+\!
G_{xz}\left(\mathbf{r}_r, \mathbf{r}^{\prime}\right) E^z_{out}\left(\mathbf{r}^{\prime}\right)
\right]
d \mathbf{r}^{\prime}. 
\label{3dee}%
\end{align}   
}%
\textcolor{black}{
Substituting (\ref{e1})-(\ref{e3}) into (\ref{3dee}), we can reformulate $E_r^x$ as: 
\begin{align}
E_r^x &= k^2 \iiint_D   \chi\left(\mathbf{r}^{\prime}\right) d \mathbf{r}^{\prime} B(\mathbf{r}^{\prime},x,y), 
\label{3de}%
\end{align} 
where the coefficient $B(\mathbf{r}^{\prime},x,y)$ is defined as 
\begin{align}
& B(\mathbf{r}^{\prime},x,y) = - \frac{j \eta}{4 \pi k} \iint_{S_1}  J_x e^{-jkR} \nonumber\\
&\left\{  {G}_{xx} \left(\mathbf{r}_r, \mathbf{r}^{\prime}\right)\left[\frac{-1-jkR+k^2R^2}{R^3}+\frac{3+3 j k R-k^2R^2}{R^5}\left(x'-x\right)^2\right] \right. \nonumber\\
&  +  {G}_{xy} \left(\mathbf{r}_r, \mathbf{r}^{\prime}\right)\frac{3+3 j kR-k^2R^2}{R^5}\left(y'-y\right)\left(x'-x
\right) \nonumber\\
& \left. +  {G}_{xz}\left(\mathbf{r}_r, \mathbf{r}^{\prime}\right) \frac{3+3 j kR-k^2R^2}{R^5} z'\left(x'-x\right)\right\} \Gamma(x,y) dx dy.
\label{A}%
\end{align} 
After discretizing $B(\mathbf{r}^{\prime},x,y)$ over $M$ sampled points in $D$ and $\Gamma(x,y)$ over $N$ sampled points on $S_1$, the formula in (\ref{A}) can be simplified to
\begin{equation}
\textbf{b}= \textbf{Y} \textbf{p},
\label{linear2}%
\end{equation} 
where the matrix $\textbf{Y}$ represents the discretized integral kernel in (\ref{A}), while the vectors $\textbf{b}$ and $\textbf{p}$ collect $B(\mathbf{r}^{\prime},x,y)$ and $\Gamma(x,y)$ on all sampled points respectively.
}%

\section{Computational Imaging by Virtual EM Masks}
In order to recover the image of the target, \textit{i.e.}, to determine the unknown $T(x',y')$, we need to exploit the relationship between virtual EM masks and the received electric field intensity in each measurement.


\subsection{Imaging Method for a 2D Homogeneous Target}
\textcolor{black}{
Denote $|E_r^x|$ and $J_{x}' (x',y')$ in the $i$th measurement as 
$|E_{r,i}^x|$ and $J_{x,i}' (x',y')$, respectively.
}
On the basis of the second order correlation theory \cite{cor},
the unknown $T(x',y')$ can be recovered by calculating the correlation between $|E_{r,i}^x|$, $J_{x,i}' (x',y')$, and $K(x',y')$ after $I$ times sampling measurements. 
Intuitively, we define the function $G_1(x',y')$ by summing the calculated current distribution $J_{x,i}' (x',y')$ with $|E_{r,i}^x|$ as the appropriate weights, \textit{i.e.},
\begin{align}
&G_1(x',y') \overset{\Delta}{=} \frac{1}{I} \sum_{i=1}^I
\left(|E_{r,i}^x|-\langle |E_{r,i}^x|\rangle_I \right) |J_{x,i}' (x',y')| \nonumber\\
&= \frac{1}{I} \sum_{i=1}^I 
\left[ 
\left|\iint_{S_2} K(\tilde{x}',\tilde{y}') T(\tilde{x}',\tilde{y}') J_{x,i}' (\tilde{x}',\tilde{y}') d \tilde{x}' d\tilde{y}') \right| -\right.\nonumber\\
& \left.
\left\langle \left|\iint_{S_2} K(\tilde{x}',\tilde{y}') T(\tilde{x}',\tilde{y}') J_{x,i}' (\tilde{x}',\tilde{y}') d \tilde{x}' d\tilde{y}') \right|
\right\rangle_I \right] |J_{x,i}' (x',y')| .
\label{q1}
\end{align}
The unknown $T(x',y')$ can be recovered from (\ref{q1}) by solving the integral equation, which is too complicated and thus needs to be simplified.
In order to simplify the expression in (\ref{q1}),
we need to exchange the orders of integral, summation, and taking the amplitude operation, which requires the phase of the integrand being a constant in the region of integration.
\textcolor{black}{
Specifically, when the condition 
$K(\tilde{x}',\tilde{y}') J_{x,i}' (\tilde{x}',\tilde{y}') \in \mathbb{R}, \forall i$
holds true, the operation $|\cdot|$ can be omitted directly.
}
To satisfy the phase of the integrand being a constant, we should design the RIS properly such that 
\begin{align}
\angle J_{x,i}' =- \angle K(\tilde{x}',\tilde{y}')
=\frac{\pi}{2}+k R' , \quad\quad\forall i.
\label{phase}
\end{align}
\textcolor{black}{
Then the integrand in (\ref{q1}) becomes a real number,
and (\ref{q1}) reduces to (\ref{q2}) at the top of this page.}

\begin{figure*}[t]
\textcolor{black}{
\begin{align}
&G_1(x',y')= \frac{1}{I} \sum_{i=1}^I 
\iint_{S_2} |K(\tilde{x}',\tilde{y}')| T(\tilde{x}',\tilde{y}') \left( |J_{x,i}' (\tilde{x}',\tilde{y}')| - 
\left\langle  |J_{x,i}' (\tilde{x}',\tilde{y}')| 
\right\rangle_I \right) |J_{x,i}' (x',y')| d \tilde{x}' d\tilde{y}'  \nonumber\\
&=\frac{1}{I} \iint_{S_2} |K(\tilde{x}',\tilde{y}')| T(\tilde{x}',\tilde{y}') \left( \sum_{i=1}^I  |J_{x,i}' (\tilde{x}',\tilde{y}')| |J_{x,i}' (x',y')|
-  \left\langle  |J_{x,i}' (\tilde{x}',\tilde{y}') |
\right\rangle_I \sum_{i=1}^I  | J_{x,i}' (x',y') |
 \right)  d \tilde{x}' d\tilde{y}' \nonumber\\
&=\iint_{S_2} |K(\tilde{x}',\tilde{y}')| T(\tilde{x}',\tilde{y}') \left( 
\left\langle |J_{x,i}' (\tilde{x}',\tilde{y}')| |J_{x,i}' (x',y') |\right \rangle_I
-  \left\langle |J_{x,i}' (\tilde{x}',\tilde{y}')|\right \rangle_I  \left\langle |J_{x,i}' (x',y')| \right \rangle_I \right)  d \tilde{x}' d\tilde{y}' .
\label{q2}
\end{align}
\hrulefill
}%
\end{figure*}

Denote $R_0'=\left.R'\right|_{x'=0,y'=0}$. Because $x'$ and $y'$ are relatively small compared to $R_0'$, we can expand $R'$ in the vicinity of $x'=0,y'=0$ based on the first-order Taylor expansion as
\begin{align}
R' &\approx R_0'+ \left.\frac{\partial R'}{\partial x'}\right|_{x'=0,y'=0} x' + \left.\frac{\partial R'}{\partial y'}\right|_{x'=0,y'=0} y' \nonumber \\
&= R_0' - \frac{x_r}{R_0'} x' - \frac{y_r}{R_0'} y' . 
\label{talor}
\end{align}
With (\ref{talor}), (\ref{phase}) can be simplified as a linear function of $x'$ and $y'$ 
\begin{align}
\angle J_{x,i}' =\frac{\pi}{2}+k (R'_0 - \frac{x_r}{R_0'} x' - \frac{y_r}{R_0'} y' ) , \quad\forall i=1,\cdots,I.
\end{align}
In order to calculate $T(x',y')$ from (\ref{q2}), 
one needs to solve a Fredholm integral equation of the first kind, which will bring heavy computation burden and considerable numerical error.
To circumvent this problem, we manage to decouple the contribution of $T(x',y')$ in (\ref{q2}). 
One effective approach is to design the RIS properly, such that 
the weights multiplied on $T(x',y')$ become Delta function. In other words, we will design the current pattern $J_{x,i}' (x',y')$ generated by the RIS to satisfy   
\begin{align}
&\left\langle |J_{x,i}' (\tilde{x}',\tilde{y}')| |J_{x,i}' (x',y')| \right \rangle_I
-  \left\langle |J_{x,i}' (\tilde{x}',\tilde{y}')| \right \rangle_I  \left\langle |J_{x,i}' (x',y')| \right \rangle_I \nonumber \\
&= c(x',y') \delta(x'-\tilde{x}',y'-\tilde{y}'),
\label{delta}
\end{align}
\textcolor{black}{
where $c(x',y')$ is a real scalar value. Then (\ref{q2}) reduces to 
\begin{align}
G_1(x',y')= c(x',y') |K(x',y')| T(x',y').
\label{q4}
\end{align}
}%
Thus, the contributions of $T(x',y')$ on different points are decoupled, and $T(x',y')$ can be straight forwardly reconstructed by using $G_1(x',y')$, $c(x',y')$, and $K(x',y')$.
Although we successfully find a straight forward solution to avoid solving the first-kind Fredholm integral equation in (\ref{q2}),
the price is that we need to design $J_{x,i}'$ that satisfies (\ref{phase}) and (\ref{delta}) beforehand.

By substituting the definition of $G_1(x',y')$ into (\ref{q4}), the image of the target is reconstructed as 
\textcolor{black}{
\begin{align}
\hat{T}(x',y')=\frac{\sum_{i=1}^I
\left(|E_{r,i}^x|-\langle |E_{r,i}^x|\rangle_I \right) |J_{x,i}' (x',y')|}{I c(x',y') |K(x',y')| }.
\label{q5}%
\end{align}
}%
Considering the fact that  
$J_{x,i}' (x',y')$ 
is induced by different distribution of $H_{\text{out}}^y$, which actually plays the role of masks that encode the information of the target, we refer to $H_{\text{out}}^y$ as \textit{virtual EM masks} in the following context. 
Note that the demanded properties of virtual EM masks, i.e., (\ref{phase}) and (\ref{delta}), are independent from the image $T(x',y')$.
Thus the same set of virtual EM masks can be repeatedly used for different targets. 
After designing a fixed set of masks to satisfy (\ref{phase}) and (\ref{delta}), we then inversely compute $\Gamma(x,y)$ that can generate the required masks on the plane of the target.

Note that those virtual masks together with the corresponding $\Gamma(x,y)$ are all computed offline, which does not increase computation burden for image reconstruction. 
Hence, the computation complexity in imaging process only comes from computing $T(x',y')$ on $M$ different sampled points by (\ref{q5}), which is $\mathcal{O}(MI)$. However, the conventional methods to solve this inverse problem with $M$ unknowns typically have computation complexity as high as $\mathcal{O}(M^3)$ \cite{linear_algebra}.
Since $I$ and $M$ are usually of the same order of the magnitude, the proposed reconstruction method can be accomplished with relatively small computation overhead.





In practice, the ideal function of $\delta(x'-\tilde{x}',y'-\tilde{y}')$ in (\ref{delta}) is impossible to be realized. 
The approximated Delta function can be realized 
by truncating the spatial spectrum of $\delta(x'-\tilde{x}',y'-\tilde{y}')$ in a finite region.
Assume that the wavenumbers in $x$ and $y$ directions are bounded by $k_x \in\left[-k_x^m, k_x^m\right]$ and $k_y \in \left[-k_y^m, k_y^m\right]$, respectively (for target symmetrical with respect to the origin).
The spatial spectrum of $\delta(x'-\tilde{x}',y'-\tilde{y}')$ is computed by Fourier transform (FT) as 
\begin{align}
F\left(k_x, k_y\right)&=\iint_{S_2} \delta(x'-\tilde{x}',y'-\tilde{y}') e^{-j k_x x'-j k_y y'} \mathrm{~d} x' \mathrm{~d} y'\nonumber \\&=e^{-j k_x \tilde{x}'-j k_y \tilde{y}'} .
\end{align}
Then the approximation of $\delta(x'-\tilde{x}',y'-\tilde{y}')$ is computed by truncating the spatial spectrum and applying inverse Fourier transform as
\begin{align}
\tilde{f}(x', y') & =\int_{-k_y^m}^{k_y^m} \int_{-k_x^m}^{k_x^m} e^{j k_x\left(x'-\tilde{x}'\right)} e^{j k_y\left(y'-\tilde{y}'\right)} \mathrm{d} k_x \mathrm{~d} k_y 
\nonumber \\&=k_x^m k_y^m \operatorname{sinc}\left[k_x^m\left(x'-\tilde{x}'\right)\right] \operatorname{sinc}\left[k_y^m\left(y'-\tilde{y}'\right)\right],
\end{align}
where $\operatorname{sinc}\left[a\right]\overset{\Delta}{=}\sin(a)/a$.
The first zeros of the sinc functions occur at $x'-\tilde{x}'=\pi / k_x^m\overset{\Delta}{=}\delta_x$ and $y'-\tilde{y}'=\pi / k_y^m\overset{\Delta}{=}\delta_y$. 
Two different points on the target have to be separated by more than $\delta_x$ and $\delta_y$ in $x$ and $y$ directions respectively, such that their correlation function defined in (\ref{delta}) can be $0$.
Thus, $\delta_x$ and $\delta_y$ define the cross-range resolutions in $x$ and $y$ directions, respectively.

Moreover, $k_x^m$ and $k_y^m$ are determined based on the  reflection path from RIS to the target. 
Denote $\theta^R_x$ and $\theta^R_y$ as the angles subtended by the RIS  along $x$ and  $y$ directions, respectively. 
Since the size of the target is relatively small compared to the RIS, 
$\theta^R_x$ and $\theta^R_y$ are regarded as two constants for any point on the target. 
Let us first show how to compute $k_x^m$.
The maximum phase shift between $x'$ and $\tilde{x}'$ occurs when the EM waves travel from the edge of the RIS and equals $k(x'-\tilde{x}')\sin(\theta^R_x/2)$ under the condition $x'-\tilde{x}'\ll z'$. Therefore, $k_x^m=k (x'-\tilde{x}')\sin(\theta^R_x/2)/(x'-\tilde{x}')=k\sin(\theta^R_x/2)$ and the cross-range resolution is
\begin{align}
\delta_x=\frac{\pi}{k_x^m}=\frac{\pi}{k \sin(\theta^R_x/2)}=\frac{\lambda}{2\sin(\theta^R_x/2)}.
\label{reso}
\end{align}
Since the center of the target is located at $(0,0,z')$, there is
\begin{align}
\sin(\theta^R_x/2)=\sqrt{\frac{(a/2)^2}{(a/2)^2+z'^2}}=\frac{a}{\sqrt{a^2+4z'^2}}.
\end{align}
The discussion for the resolution in the $y$ direction is similar.

Since the images are reconstructed using a weighted sum of the masks, the cross-range 
resolution of the images is the same as the cross-range 
resolution of the masks.
Note that the resolution is irrelevant to the location of the receiver.
Even though the receiver is located in the far field of the target,
the resolution less than the wave length can still be achieved, which is contrary to the common sense.
Physically speaking, the incident and the scattered waves on the target both consist of
evanescent waves and propagating waves.
The evanescent waves carry the high-spatial-frequency components, while the propagating waves carry the low-spatial-frequency components.
Upon scattering on the target, conversions take place from evanescent-wave spectrum into propagating-wave spectrum. 
The high-spatial-frequency components in the evanescent spectrum of the virtual EM masks are encoded into propagating spectrum of the scattered waves and thus can reach the far field.
Therefore, the far-field receiver can sense the high-spatial-frequency components of the virtual EM masks and  thus achieve resolution less than the wave length.


\begin{remark}
The inspiration of reconstructing $T(x',y')$ by the cross correlation between $|E_{r,i}^x|$ and $|J_{x,i}' (x',y')|$
comes from the ghost imaging technique in optics \cite{cor,cor2,cor3,cor4,cor5}. 
However, there are two main differences between the proposed approach and the ghost imaging technique. 
First, in the conventional ghost imaging technique the random speckle patterns are measured by a charge coupled device or a scanning pinhole detector. While in the proposed approach, the speckle patterns are calculated through rigorously derived EM formulas.               
Second, the conventional ghost imaging technique needs a changeable light source such as a digital micromirror device (DMD) to generate pre-designed speckle patterns and to project the speckle patterns onto the object.
While in this paper, the EM source is fixed and unchangeable, and we propose a RIS design scheme to generate the desired masks instead. 
\end{remark}

\textcolor{black}{
\subsection{Imaging Method for a 3D Inhomogeneous Target}
When imaging a 3D inhomogeneous dielectric target, both the amplitude and the phase of $E_r^x$ are indispensable.
Denote $E_r^x$ and $B(\mathbf{r}^{\prime})$ in the $i$th measurement as 
$E_{r,i}^x$ and $B_i(\mathbf{r}^{\prime})$, respectively. 
Similar to the 2D imaging method,
we define the function $G_2(\mathbf{r}^{\prime})$ as 
\begin{align}
&G_2(\mathbf{r}^{\prime})\overset{\Delta}{=} \frac{1}{I} \sum_{i=1}^I
\left(E_{r,i}^x-\langle E_{r,i}^x \rangle_I \right) B_i(\mathbf{r}^{\prime}) \nonumber\\
& = k^2 \iiint_D   \chi\left(\tilde{\mathbf{r}}^{\prime}\right) d\tilde{\mathbf{r}}^{\prime} \left[ \left\langle B_i(\tilde{\mathbf{r}}^{\prime})
B_i(\mathbf{r}^{\prime})\right\rangle_I \right. - \left.
\left\langle  B_i(\mathbf{r}^{\prime})
\right\rangle_I  
\left\langle  B_i(\tilde{\mathbf{r}}^{\prime})
\right\rangle_I
\right] .
\label{q_2} %
\end{align}
}%
\textcolor{black}{
In this case, $B_i(\tilde{\mathbf{r}}^{\prime})$ plays the role of the virtual masks that should satisfy 
\begin{align}
&\left\langle B_i(\tilde{\mathbf{r}}^{\prime}) B_i(\mathbf{r}^{\prime}) \right \rangle_I
-  \left\langle B_i(\tilde{\mathbf{r}}^{\prime}) \right\rangle_I  \left\langle B_i(\mathbf{r}^{\prime}) \right \rangle_I \nonumber \\
&= c(\mathbf{r}^{\prime}) \delta(\mathbf{r}^{\prime}-\tilde{\mathbf{r}}^{\prime}, \mathbf{r}^{\prime}-\tilde{\mathbf{r}}^{\prime}),
\label{delta_2}%
\end{align}
where $c(\mathbf{r}^{\prime})$ is an arbitrary scalar value. Then (\ref{q_2}) reduces to 
\begin{align}
G_2(\mathbf{r}') = k^2 c(\mathbf{r}') \chi\left(\mathbf{r}^{\prime}\right).
\label{q4_2}
\end{align}
Thus, the image of the target is reconstructed as 
\begin{align}
\hat{\chi}(\mathbf{r}^{\prime}) = \frac{\sum_{i=1}^I
\left(E_{r,i}^x-\langle E_{r,i}^x\rangle_I \right) B_{i} (\mathbf{r}^{\prime})}{I k^2 c(\mathbf{r}') }.
\label{q52}%
\end{align}
}%

\section{Design and Generation of Virtual EM Masks}
The holographic active RIS can provide massive degrees of freedom to modulate the phase, amplitude, polarization together with multiple parameters of EM waves within sub-wavelength resolution \cite{h1,h2,active1,active2}. 
Thanks to these properties, we can generate virtual masks by properly designing the holographic active RIS to satisfy the reconstruction prerequisites, i.e., (\ref{phase}) and (\ref{delta}).

\subsection{Design of 2D Virtual EM Masks}
For the $i$th measurement, let $\textbf{q}_i= [q_{i,1},\cdots,q_{i,M}]^T \in \mathbb{R}^M$ and $\textbf{g}_i \in \mathbb{R}^M$ collect amplitude and phase of $J_{x,i}' (x',y')$ over $M$ sampled points on $S_2$, respectively.
\textcolor{black}{
In the following, we separately design $\textbf{g}_i$ and $\textbf{q}_i$ in detail to satisfy the prerequisites of the proposed reconstruction method, which are listed in (\ref{phase}) and (\ref{delta}).
}

For different measurements, $\textbf{g}_i$ is invariant and can be directly computed according to (\ref{phase}). 
After the sampling process, the requirement for amplitude in (\ref{delta}) can be rewritten in the discretized form as 
\begin{align}
\frac{1}{I} \sum_{i=1}^I q_{i,m} q_{i,n} - \frac{1}{I^2} \sum_{i=1}^I q_{i,m} \sum_{i=1}^I q_{i,n} = C_{n,m} \delta_{nm}.
\label{delta2}
\end{align}
To satisfy (\ref{delta2}), we generate $\textbf{q}_i$ by using the partial columns of the Hadamard matrix $\textbf{H}$. 
Each column of $\textbf{H}$ is utilized to generate the mask value on a corresponding target point.
Hence, the order of the Hadamard matrix $\textbf{H}$ should be equal to the number of measurements $I$, while the number of the partial columns should be equal to the number of the sampled points $M$. 
Since the first column of $\textbf{H}$ is only composed of $1$, using the first column to generate $\textbf{q}_i$ will make $J_{x}' (x',y')$ a constant on the first sampled point, which provides no information of the scattering density at this point. 
Thus, we abandon the first column of $\textbf{H}$ and assign the $(i,m+1)$th entry of $\textbf{H}$ to $q_{i,m}$ ($m=1,\cdots,M$). According to the property of Hadamard matrix, there are
\begin{align}
\sum_{i=1}^I q_{i,m} q_{i,n}&=I \delta_{nm}, 
\label{p1} \\
\sum_{i=1}^I q_{i,m} &= 0.
\label{p2}
\end{align}
Because of (\ref{p1}) and (\ref{p2}), (\ref{delta2}) can be satisfied with $C_{n,m}=1$.
Moreover, because the Hadamard matrix $\textbf{H}$ is a square matrix whose order can only be an integer multiple of $4$, $I$ must be chosen as an integer multiple of $4$, too.

Since the elements of $\textbf{q}_i$ are amplitudes of $J_{x,i}' (x',y')$ that should be non-negative values, we transform the elements of $\textbf{q}_i$ from $-1$ to $0$.
In other words, the magnetic fields are supposed to interfere destructive at points where $q_{i,m}=0$.
Apparently, the transformation of $q_{i,m}$ from $-1$ to $0$ is equivalent
to $(1+q_{i,m})/2$, which does not influence the orthogonal property of (\ref{delta2}) with $C_{i,m}=1/4$ after the transformation.

\textcolor{black}{
It is essential that the information provided by each $\textbf{q}_i$ should be as different as possible, which places demands on the prerequisite in (\ref{p1}). 
The property of (\ref{p1}) is defined as the \textit{orthogonality} between masks \cite{orthogonal}.
Generally speaking, the sharper orthogonality would yield more information of the target. 
Note that although we use Hadamard masks as an example, $\textbf{q}_i$ can also be designed using other patterns satisfying (\ref{p1}), such as Gaussian random patterns and Fourier patterns. 
}

\subsection{Generating 2D Virtual EM Masks by the RIS}
\begin{figure}[t]
  \centering  
\centerline{\includegraphics[width=7.4cm,height=5.5cm]{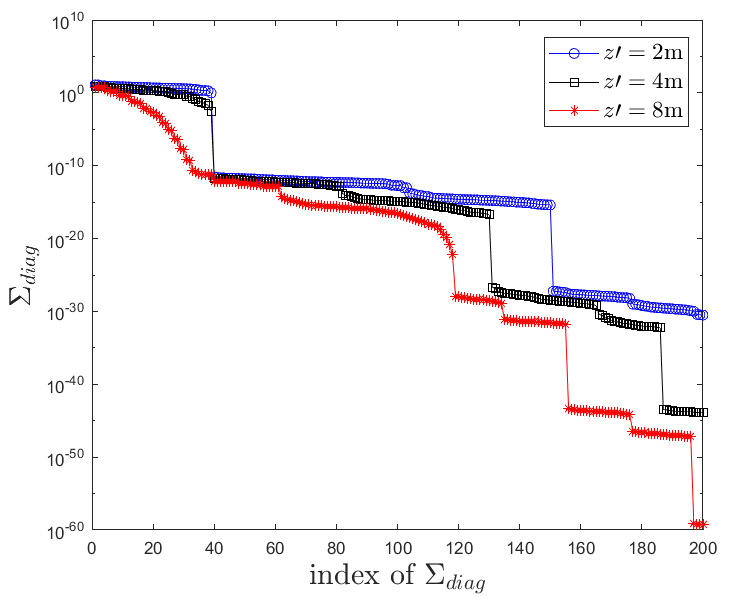}}
  \caption{Singular value spectrum of $\textbf{Z}$}
  \label{SVS}
\end{figure}
According to (\ref{twice}), $J_{x,i}' (x',y')$ is equal to $(1-\Gamma')H_{\text{out}}^y$, where $(1-\Gamma')$ is assumed to be a constant over the target composed of homogeneous material.
Thus, we aim to control the distribution of $H_{\text{out}}^y$ by designing $\textbf{p}$ in (\ref{linear}). 
Note that the ignorance of the specific value of $\Gamma'$ does not influence the
synthesis of EM fields.

Denote $\textbf{p}_i$ and $\textbf{y}_i$ as $\textbf{p}$ and $\textbf{y}$ in the $i$th measurement, respectively.
With the knowledge of $\textbf{y}_i$ and $\textbf{Z}$, we can acquire the corresponding $\textbf{p}_i$ by solving the linear equations in (\ref{linear}). 
The calculation of $\textbf{p}_i$ by using conventional least squares techniques, however, can become ill-conditioned as $N$ and $M$ grow relatively large.
To circumvent this ill-conditioned problem, we may reformulate the least squares problem by applying Tikhonov regularization as
\begin{equation}
\min _{\boldsymbol{\textbf{p}_i}}\left(\|\mathbf{Z} \boldsymbol{\textbf{p}_i}-\textbf{y}_i\|_2^2+\gamma\|\boldsymbol{\textbf{p}_i}\|_2^2\right) ,
\label{tik}
\end{equation}
where $\gamma$ is a positive regularization parameter that decides the weight of regularization term in (\ref{tik}).


In practical implementation, (\ref{tik}) can be solved by constructing another matrix, $\textbf{Z}^{\mathrm{Tik}}$, which acts as a
pseudo-inverse operator. 
Denote singular value decomposition (SVD) of $\textbf{Z}$ as  $\textbf{Z}=\textbf{U}\boldsymbol{\Sigma}\textbf{V}^H$.
The key  of Tikhonov regularization is to find the inverse of $\boldsymbol{\Sigma}$ modified by the regularization term in (\ref{tik}).
Let $\boldsymbol{\Lambda} \in \mathbb{C}^{N\times M}$ denote the regularized pseudo-inverse of $\boldsymbol{\Sigma}$
and let $\Sigma_{\text {diag}}$ denote the diagonal components of $\boldsymbol{\Sigma}$. 
Then the diagonal elements of $\boldsymbol{\Lambda}$ can be computed as 
\begin{equation}
\Lambda_{\text {diag }}=\frac{\Sigma_{\text {diag}}}{\Sigma_{\text {diag }}^2+\gamma}.
\end{equation}
Next, the regularized pseudo-inverse of $\textbf{Z}$ is computed as
\begin{equation}
\textbf{Z}^{\mathrm{Tik}}=\textbf{V}\boldsymbol{\Lambda}\textbf{U}^H.
\label{tik2}
\end{equation}

When we compute $\boldsymbol{\Sigma}$ as a regularized inverse of $\boldsymbol{\Lambda}$,
the singular values much larger than $\gamma$ do not have any significant modification,
whereas the singular values much smaller than $\gamma$ make negligible contribution to the solution of $\textbf{Z}^{\mathrm{Tik}}$.
Thus, $\Lambda_{\text {diag}}$ can be set directly as $0$ when $\Sigma_{\text {diag}}$ drops to a certain threshold that is much less than $\gamma$, e.g., $10^{-5}\gamma$.
When the relatively small $\Lambda_{\text {diag}}$ are removed, we only need to utilize the major current patterns on the RIS to generate the desired EM masks. 
Therefore, the rank of $\textbf{Z}^{\mathrm{Tik}}$ is reduced, and the
computational efficiency is improved. 
\begin{figure*}[t]
\begin{minipage}[t]{0.5\linewidth}
\centering
\subfigure[the phase of reflecting coefficients on the RIS]{
\includegraphics[width=6.5cm,height=5.5cm]{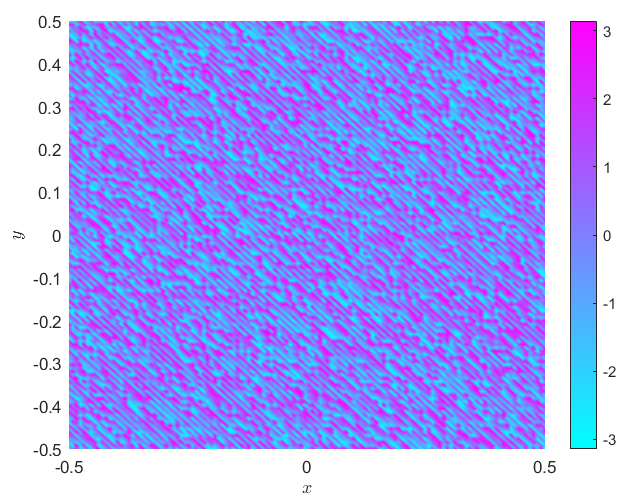}}
\end{minipage}
\begin{minipage}[t]{0.5\linewidth}
\centering
\subfigure[the amplitude of reflecting coefficients on the RIS]{
\includegraphics[width=6.5cm,height=5.5cm]{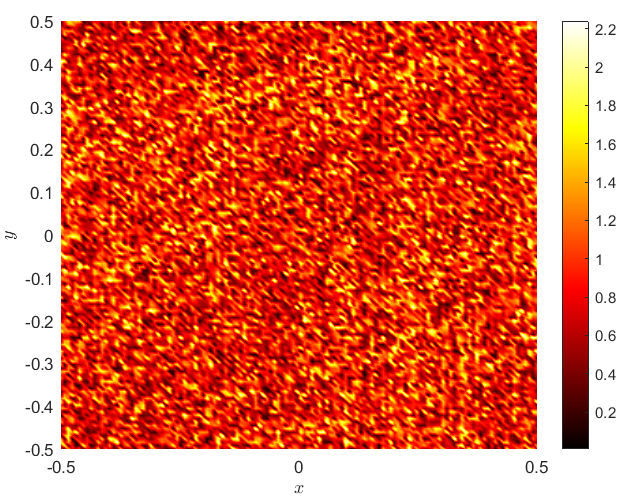}}
\end{minipage}
\caption{
Phase and amplitude of the coefficients on the RIS when generating one of the virtual masks, where we choose $N=16384$, $P_I=1$.
}
\label{tau}
\end{figure*}

To provide a visual example of the distribution of singular values, 
we show the values of the $200$ largest $\Sigma_{\text{diag}}$ in Fig.~\ref{SVS}
when $M=4096$ and $N=16384$. 
It is seen that $\Sigma_{\text {diag}}$ decays rapidly in a staircase curve with the growth of singular value index. 
The sharp decay of the singular values leads to an enormous condition number of $\textbf{Z}$, which makes it necessary to apply Tikhonov regularization.
The well-coupled modes for RIS and target planes correspond to those large singular values, whose number is small compared to $M$ and $N$.
When $z'$ becomes larger, $\Sigma_{\text {diag }}$ decays more sharply, and the number of dominant singular values is smaller.

After computing $\textbf{Z}^{\mathrm{Tik}}$ by (\ref{tik2}), $\boldsymbol{\textbf{p}_i}$ can be obtained  directly according to the Moore-Penrose solution as
\begin{equation}
\tilde{\textbf{p}}_i=\mathbf{Z}^{\mathrm{Tik}} \textbf{y}_i.
\end{equation}
Considering the total amplification rate $P_I$ at RIS, the implemented reflection coefficient should be normalized as 
\begin{equation}
\textbf{p}_i=\sqrt{N P_I} \frac{\tilde{\boldsymbol{\textbf{p}}}_i}{\Vert\tilde{\boldsymbol{\textbf{p}}}_i\Vert}.
\label{final_p}
\end{equation}

\begin{figure*}[t]
\begin{minipage}[t]{0.32\linewidth}
\subfigure[ phase of ideal mask]{
\includegraphics[width=5.5cm,height=5.0cm]{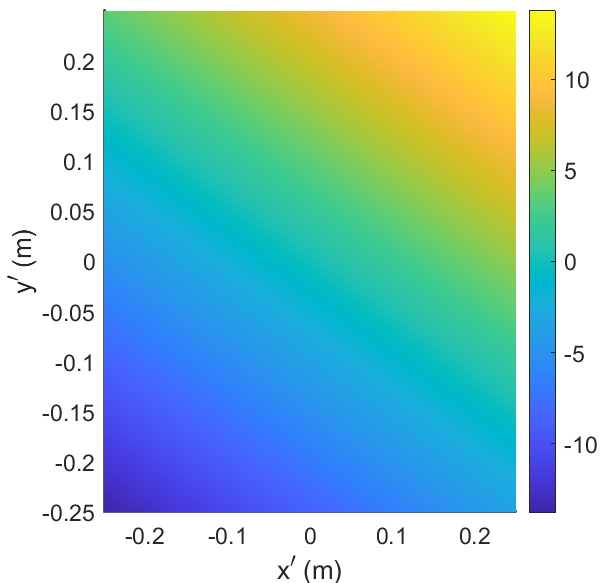}}
\end{minipage}
\begin{minipage}[t]{0.32\linewidth}
\subfigure[ phase of generated mask, $z'=2$~m]{
\includegraphics[width=5.5cm,height=5.2cm]{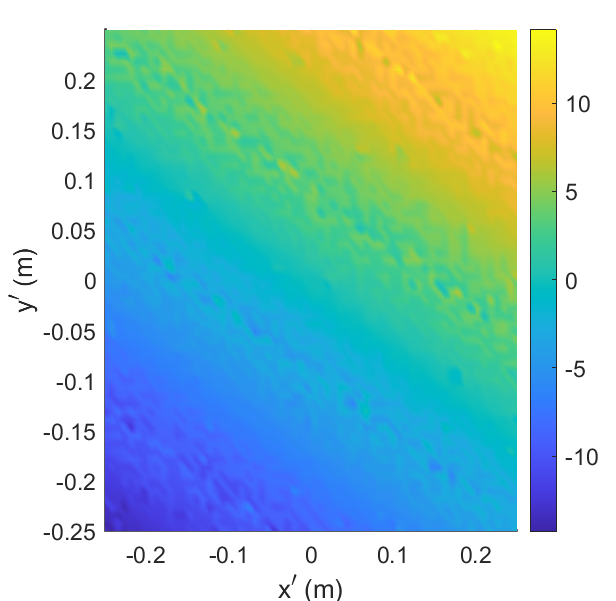}}
\end{minipage}
\begin{minipage}[t]{0.32\linewidth}
\subfigure[phase of generated mask, $z'=8$~m]{
\includegraphics[width=5.5cm,height=5.0cm]{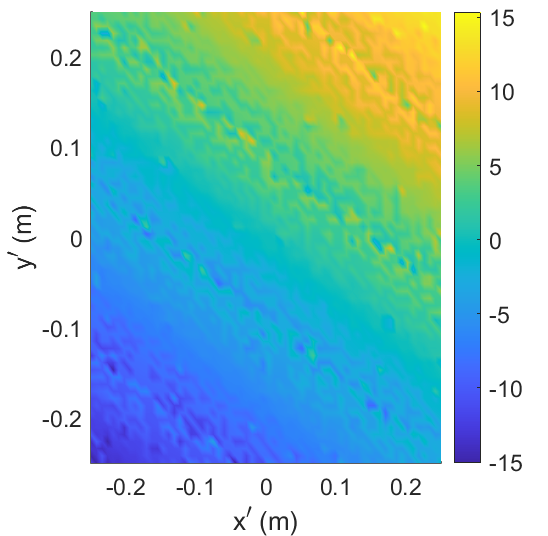}} 
\end{minipage}  \\
\begin{minipage}[t]{0.32\linewidth}
\subfigure[ amplitude of ideal mask]{
\includegraphics[width=5.5cm,height=5.0cm]{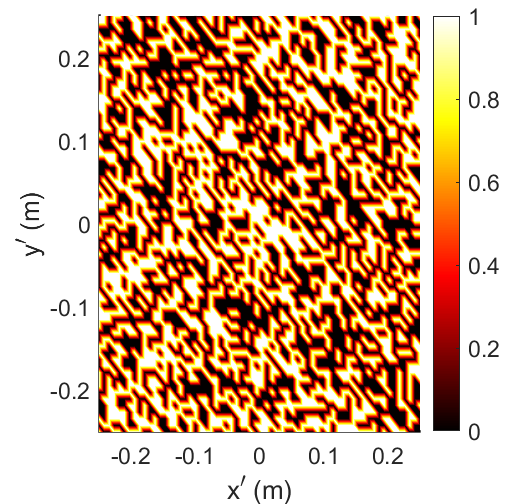}}
\end{minipage}
\begin{minipage}[t]{0.32\linewidth}
\subfigure[ amplitude of generated mask, $z'=2$~m]{
\includegraphics[width=5.5cm,height=5.0cm]{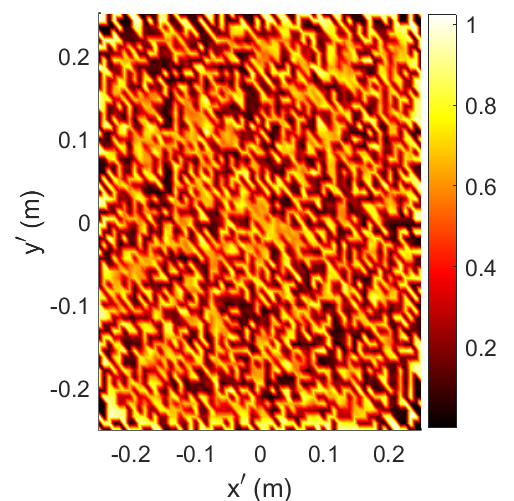}}
\end{minipage}
\begin{minipage}[t]{0.32\linewidth}
\subfigure[ amplitude of generated mask, $z'=8$~m]{
\includegraphics[width=5.5cm,height=5.0cm]{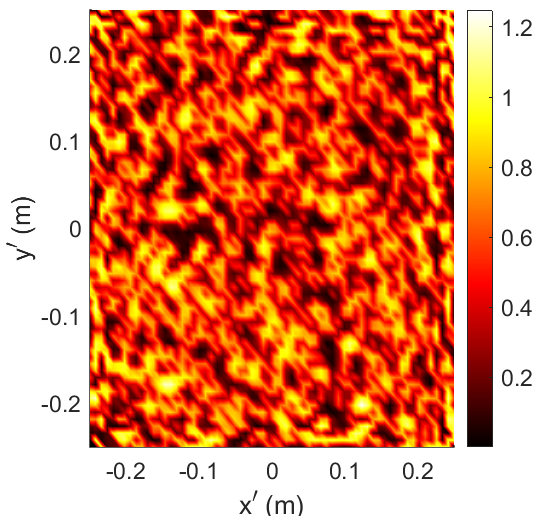}}
\end{minipage}
\caption{
\textcolor{black}{Comparison between the ideal mask and the generated mask.
The degree of brightness represents the strength of the magnetic field in $y$ direction.
The maximum amplitude in each mask is normalized to $1$.}
}
\label{hada}
\end{figure*}

In order to provide a visual example of the 2D virtual EM mask generated by RIS, we show the phase and the amplitude of the RIS in Fig.~\ref{tau} and compare the ideal mask with the generated mask in Fig.~\ref{hada}.
The system parameters are set as  $P_I=1$, $x_r=15$~m, $y_r=25$~m, $z_r=-30$~m, $N=16384$, $M=4096$, $\lambda=0.01$~m, $a=b=1$~m, and $a'=b'=0.5$~m.
\textcolor{black}{
The Tikhonov regularization parameter is set as $\gamma=10^{-12}$ and $\gamma=10^{-15}$ for $z' = 2$~m and $z' = 8$~m, respectively.
}
It is seen from Fig.~\ref{tau} that both the phase and the amplitude of the RIS vary sharply in order to generate a complicated pattern of the virtual mask. 
The points with $|\Gamma(x,y)|<1$ represent that there is power flow from the incident EM waves into the RIS, while the points with $|\Gamma(x,y)|>1$ represent that there is power flow from the RIS to the reflected EM waves.


It is seen from Fig.~\ref{hada} that, there are differences between the ideal mask and the generated mask, while such differences become more significant with the increase of $z'$ because the EM fields are harder to manipulate in the farther zone. 
As shown in Fig.~\ref{hada}(a)-Fig.~\ref{hada}(c),
the phases of the ideal virtual mask change almost linearly with respect to $x'$ and $y'$, which verifies the approximation made in (\ref{talor}).
Although the phases of the generate virtual masks have some local differences compared with Fig.~\ref{hada}(a), the linear variation trend can still be observed in both Fig.~\ref{hada}(b) and Fig.~\ref{hada}(c).
Despite the amplitude discrepancy, the patterns in Fig.~\ref{hada}(e) are almost the same as those in Fig.~\ref{hada}(d) but have the difference that the image contrast in Fig.~\ref{hada}(e) is less than the sharp 0-1 contrast in the ideal Hadamard mask.
The lower contrast leads to a smaller $C_{m,n}$ in (\ref{delta2}), while it does not severely weaken the orthogonality among the virtual masks.
When $z'=8$~m, even the patterns in Fig.~\ref{hada}(f) are also distorted.
The mismatch between the ideal mask and the generated mask will degrade the orthogonality among masks, thereby reducing the imaging quality.

Since the errors are caused by the finite-sized RIS and finite sampled points on the holograghic RIS, the quality of generated field can be improved by increasing the 
overall size of RIS, which however, is limited by the manufacturing capability. 
A more practical approach to enhance the orthogonality among the masks is to decrease $z'$.

\begin{figure*}[t]
\begin{minipage}[t]{0.5\linewidth}
\centering
\subfigure[\textcolor{black}{The correlation function with respect to the central point of the target with $z'=2$~m}]{
\includegraphics[width=6.5cm,height=5.0cm]{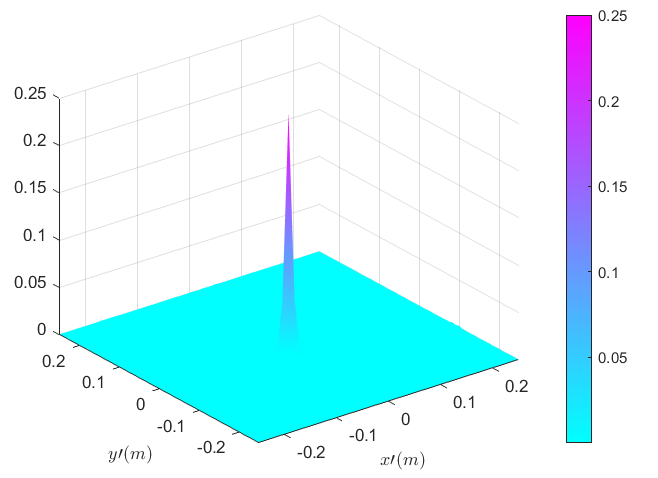}}
\end{minipage}
\begin{minipage}[t]{0.5\linewidth}
\centering
\subfigure[\textcolor{black}{The correlation function with respect to the central point of the target with $z'=8$~m}]{
\includegraphics[width=6.5cm,height=5.0cm]{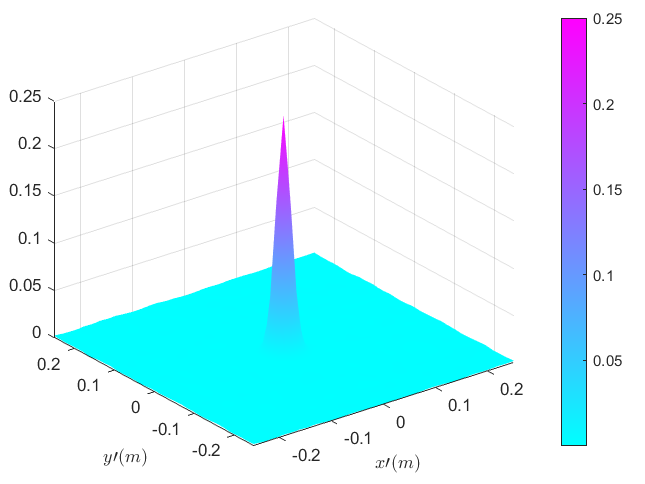}}
\end{minipage}
\caption{
Correlation functions with respect to the central point of the target are shown, where we choose $N=16384$, $M=4096$, $a'=b'=0.5$~m
}
\label{cor}
\end{figure*}

In order to show the orthogonality property among the generated masks, we calculate the correlation function with respect to the central point of the target
according to the left-hand side of (\ref{delta2}) in Fig.~\ref{cor}, with the same set of parameters in Fig.~\ref{tau} and Fig.~\ref{hada}.
It is seen from Fig.~\ref{cor} that the correlation function is not a strict Delta function due to the imperfect generated masks.
However, the correlation function is still close to $0$ except for a small region around the central point. For $z'=2$~m, the non-zero regions are more concentrated than those for $z'=8$~m, and the correlation function is more similar to Delta function.
This is because when $z'$ is smaller, $\theta^R_x$ and $\theta^R_y$ will be larger, leading to a smaller cross-range resolution $\delta_x$ and $\delta_y$ according to (\ref{reso}).

\textcolor{black}{
\subsection{Design and Generation of 3D Virtual EM Masks}
Define $\textbf{b}_i$ as the 3D virtual EM mask vector $\textbf{b}$ in (\ref{linear2}) from the $i$th measurement. 
Similar to the design of $\textbf{q}_i$ in (\ref{delta2})-(\ref{p2}), $\textbf{b}_i$ is generated by using $M$ columns of the Hadamard matrix $\textbf{H} \in \mathbb{R}^{I \times I}$. 
Each column of $\textbf{H}$ is utilized to generate the mask values on a corresponding sampled point such that (\ref{q52}) can be satisfied.
Denote $\mathbf{Y}^{\mathrm{Tik}}$ as the regularized Moore-Penrose pseudo inverse of $\mathbf{Y}$.
We can calculate $\textbf{p}_i$ as 
\begin{equation}
\textbf{p}_i=\sqrt{N P_I} \frac{\mathbf{Y}^{\mathrm{Tik}} \textbf{b}_i}{\Vert\mathbf{Y}^{\mathrm{Tik}} \textbf{b}_i\Vert}.
\end{equation}
}


\section{Simulation Results and Analysis}
\textcolor{black}{
Since the holographic RIS is still a type of hardware under study and several key challenges in manufacturing process remain to be solved \cite{infeasible}, we are temporarily unable to conduct experimental measurements. 
Instead, we use the 3D EM simulation software Ansys HFSS to simulate the imaging process and to 
generate the received signals, which are used in reconstructing the image of the object. 
}
In the simulations, we consider a square RIS of size $a = b = 2$~m.
The total amplification rate at RIS is set as $P_I = 1$.
The elevation angle of the incident EM waves on the RIS is $\theta_{\text{in}} =30\degree$.
The wavelength is $\lambda = 0.01$~m, i.e., the frequency is $29.9$ GHz.
The size of the target is $a' = b' = 0.5$~m, and the distance between the target and the RIS is adjusted in the region $z'\in [2,8]$~m. 
Correspondingly, there is $\sin(\theta^R_x/2),\sin(\theta^R_y/2) \in [0.7071,0.2425]$ and the cross-range resolution is computed from (\ref{reso}) as $\delta_x,\delta_y\in [0.0071,0.0206]$~m.
We set the location of the receiver as $(x_r,y_r,z_r)=(40,40,-10)$~m, which guarantees that the line-of-sight path of waves reflected from the target to the receiver will not be blocked by the RIS, and the receiver is located within the far field of the target, i.e., $d_r>\frac{2 (a'^2+b'^2)}{\lambda}=50$~m.

The source of noise in the simulation is thermal noise on the receiver load. We consider a system that uses a bandwidth $B=1$ $\mathrm{MHz}$, while the power spectral density of the thermal noise is $N_0=-174$ $\mathrm{dBm} / \mathrm{Hz}$. Consequently, the thermal noise power is $\sigma_n^2=N_0 B=-114 $ $\mathrm{dBm}$ and the SNR at the receiver is defined as SNR $=P_r/\sigma_n^2$.
\textcolor{black}{
Let the vector $\textbf{t} \in \mathbb{C}^M$ 
collect $T(x', y')$ or $\chi(\mathbf{r}^{\prime})$ on the $M$ sampled points for the 2D or the 3D target respectively. 
We then introduce the normalized mean square error (NMSE) as the criterion to quantitatively describe the quality of imaging performance: 
\begin{equation}
\mathrm{NMSE}=\frac{\left\|\mathbf{t}-\hat{\mathbf{t}}\right\|_{F}^{2}}{\left\|\mathbf{t}\right\|_{F}^{2}} .
\end{equation}
}


\textcolor{black}{
\subsection{Imaging Performance for a 2D Homogeneous Target}
}
The material of the 2D target is set as PEC, i.e., $\Gamma'=-1$. 
We choose the number of sampled points as $N=128 \times 128=16384$ and $M=64\times 64=4096$, respectively. 
\textcolor{black}{
The Tikhonov regularization parameter is set as $\gamma=10^{-12}$, $10^{-14}$, and $10^{-15}$ for $z' \in [2,3]$, $(3,5]$, and $(5,8]$~m, respectively.
}
Each sampled point on the target can be regarded as a pixel in the image, i.e., the picture is composed of $M=64\times 64=4096$ pixels.
The interval of the adjacent pixels is thus $a'/64=b'/64=0.0078$~m, which is of the same order of magnitude as $\delta_x$ and $\delta_y$.
More pixels could hardly improve the imaging quality due to the resolution limit.
\subsubsection{Imaging Performance versus the Number of Measurements}

\begin{figure*}[t]
  \centering
\begin{minipage}[t]{0.32\linewidth}
\subfigure[the abbreviation for Southeast University]{
\includegraphics[width=5.5cm]{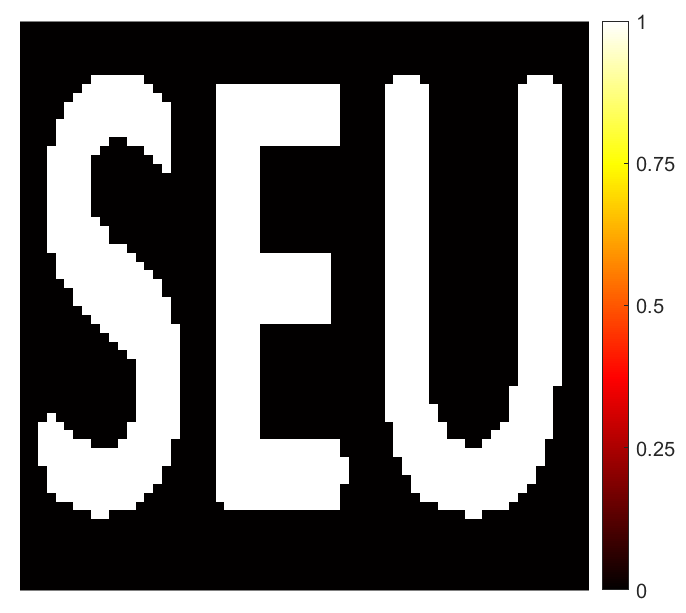}} 
\end{minipage}
\begin{minipage}[t]{0.32\linewidth}
\subfigure[the image with 4096 measurements]{
\includegraphics[width=5.5cm]{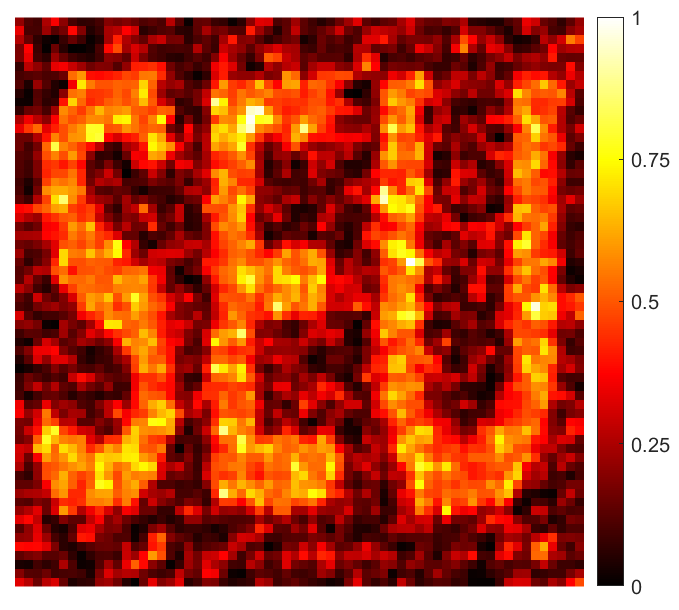}}  
\end{minipage}
\begin{minipage}[t]{0.32\linewidth}
\subfigure[the image with 16384 measurements]{
\includegraphics[width=5.5cm]{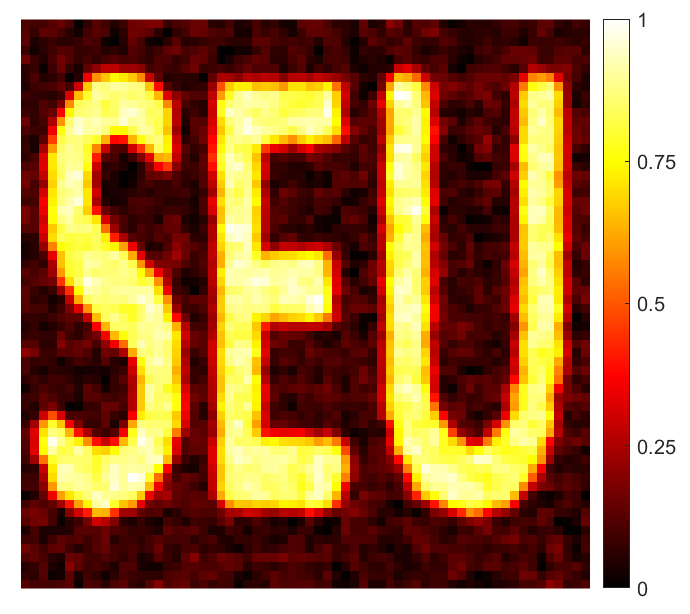}}   
\end{minipage} \\
\begin{minipage}[t]{0.32\linewidth}
\subfigure[the badge for Tsinghua University]{
\includegraphics[width=5.5cm]{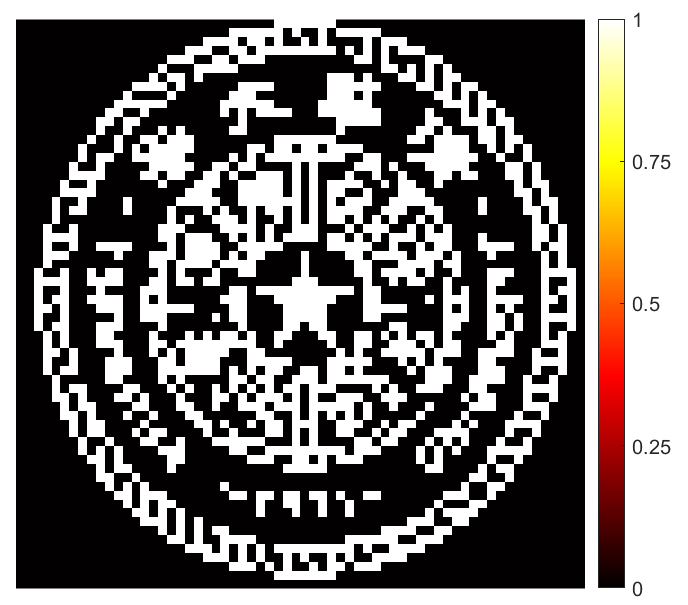}}
\end{minipage}
\begin{minipage}[t]{0.32\linewidth}
\subfigure[the image with 4096 measurements]{
\includegraphics[width=5.5cm]{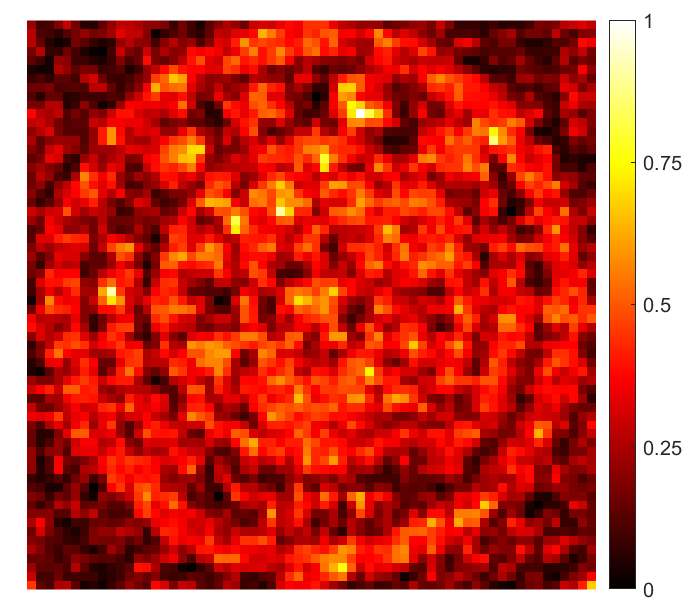}}
\end{minipage}
\begin{minipage}[t]{0.32\linewidth}
\subfigure[the image with 16384 measurements]{
\includegraphics[width=5.5cm]{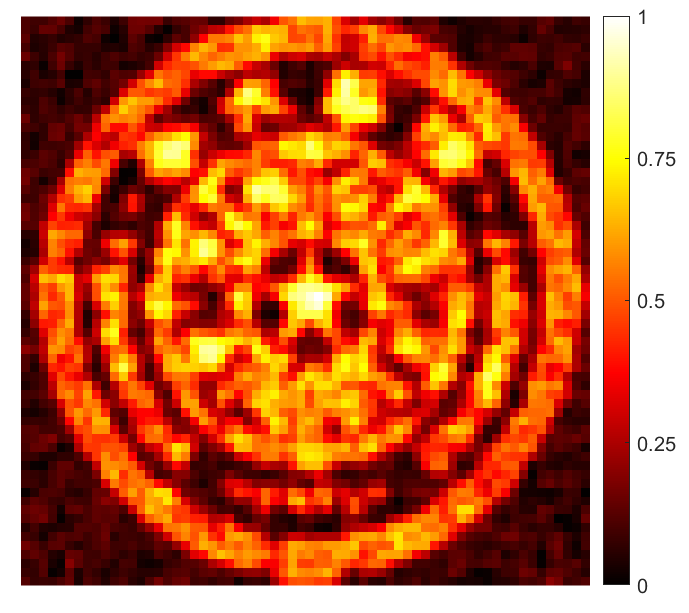}} 
\end{minipage}
\caption{\textcolor{black}{Imaging results versus the number of measurements $I$, when we set $z'=4$~m and SNR $=20$~dB.}
}
\label{image1}
\end{figure*}

\begin{figure}[t]
  \centering
\centerline{\includegraphics[width=7.4cm,height=5.5cm]{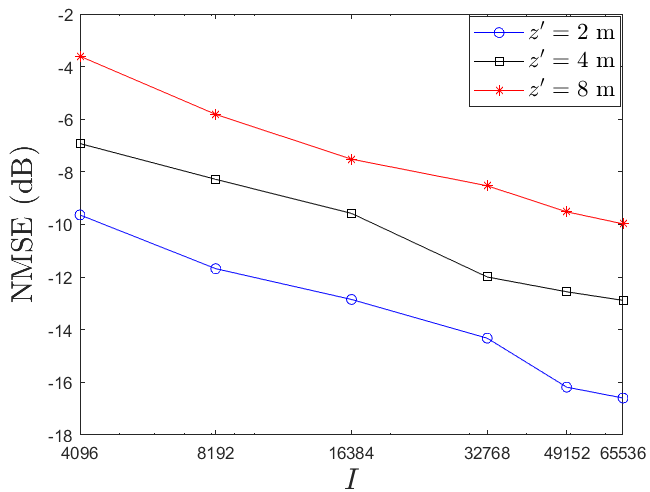}}
  \caption{\textcolor{black}{NMSE versus the number of measurements $I$, with SNR $=20$~dB}}
  \label{nmse1}
\end{figure}


We explore the NMSE versus the number of measurements $I$, as shown in Fig.~\ref{nmse1}. 
\textcolor{black}{
We set SNR at the receiver as $20$~dB and the distance of the target as $z' = 2, 4,$ and $8$~m. 
The target ``SEU" shown in Fig.~\ref{image1}(a) is used in test.}
The generations of Hadamard matrices for different $I$ are accomplished by Sylvester's construction, Paley construction, or the Turyn construction.


To illustrate the imaging result vividly, we present the reconstructed images of two target patterns when $z'=4$~m and SNR $=20$~dB.
The first target, ``SEU", is the abbreviation for Southeast University, and the second target is the school badge of Tsinghua University.
The first target pattern is simple, with more low-frequency components, while the second target pattern is complex, with more high-frequency components.
As shown in Fig.~\ref{image1}, all reconstructed images reproduce the general shapes of the targets accurately.
If $I$ is equal to the total number of pixels, i.e., $I=4096$, then the restored image will have obvious speckles, making the images blurry. 
However, when $I$ is $4$ times more than the total number of pixels, i.e., $I=16384$, the imaging result is clear, with only the background not matching the original target.
It is then evident that the reconstruction algorithm works well for a large number of measurements that compensate for the distortion of the generated masks.

It is seen from Fig.~\ref{nmse1} that, as $I$ increases, the NMSE will decrease, indicating that a larger number of measurements improves the quality of the imaging system.
When $I$ is small, the measurements are sparse, and there is insufficient information to accurately reconstruct the object.
As $I$ increases, more information of the target will be captured, and the reconstructed image becomes more accurate. 
Additionally, when the distance between the target and the RIS is reduced to $2$~m, the NMSE becomes smaller than that at the distances of $4$~m and $8$~m, indicating that the imaging performance is superior for targets in closer proximity to the RIS.

\subsubsection{Imaging Performance versus SNR}
\begin{figure*}[t]
  \centering
\begin{minipage}[t]{0.32\linewidth}
\subfigure[the abbreviation for Southeast University]{
\includegraphics[width=5.5cm]{target3.png}} 
\end{minipage}
\begin{minipage}[t]{0.32\linewidth}
\subfigure[the imaging result with SNR = 10 dB]{
\includegraphics[width=5.5cm]{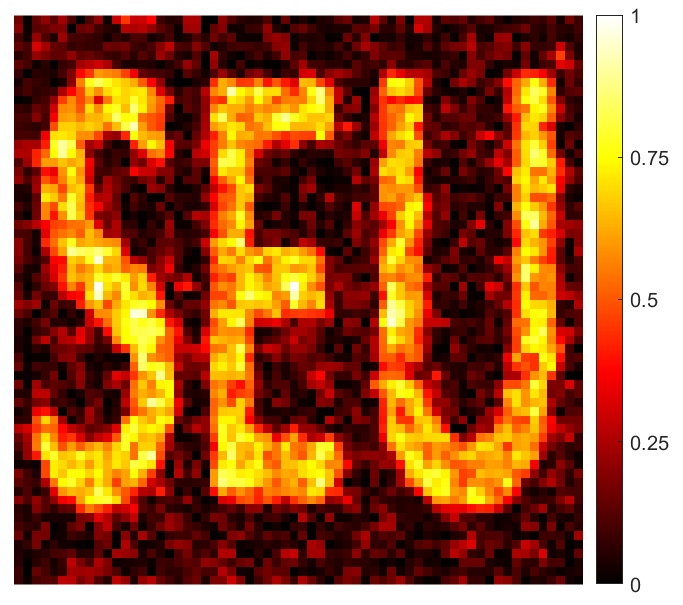}} 
\end{minipage}
\begin{minipage}[t]{0.32\linewidth}
\subfigure[the imaging result with SNR = 30 dB]{
\includegraphics[width=5.5cm]{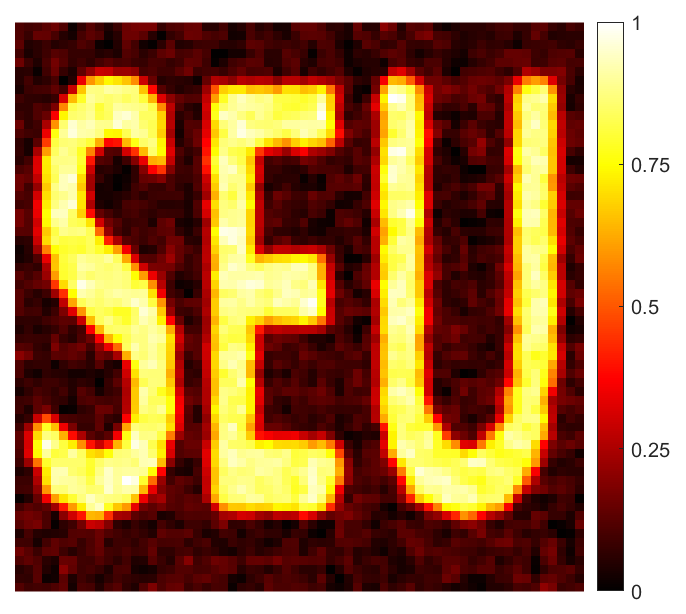}} 
\end{minipage} \\
\begin{minipage}[t]{0.32\linewidth}
\subfigure[the badge for Tsinghua University]{
\includegraphics[width=5.5cm]{target2.png}}
\end{minipage}
\begin{minipage}[t]{0.32\linewidth}
\subfigure[the imaging result with SNR = 10 dB]{
\includegraphics[width=5.5cm]{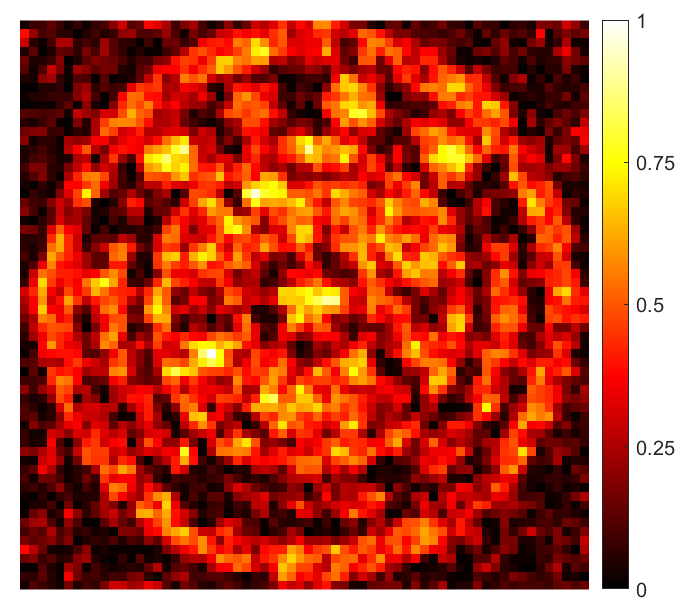}}
\end{minipage}
\begin{minipage}[t]{0.32\linewidth}
\subfigure[the imaging result with SNR = 30 dB]{
\includegraphics[width=5.5cm]{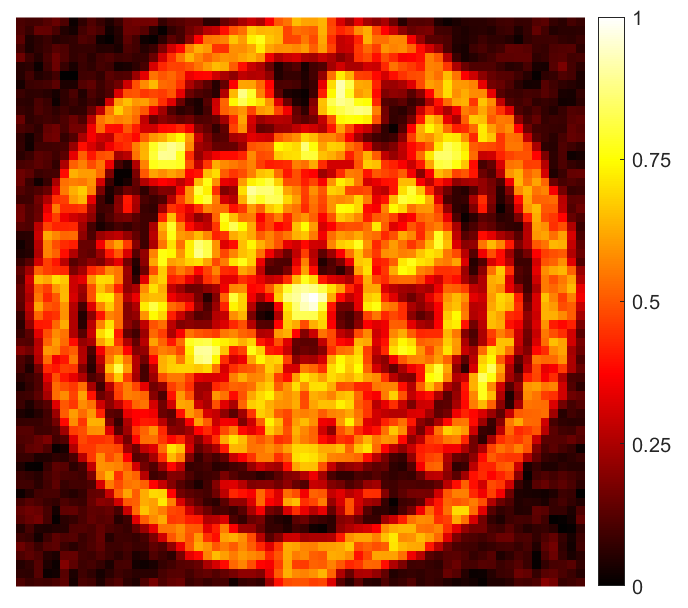}}
\end{minipage}
\caption{\textcolor{black}{Imaging results versus SNR when we set $z'=4$~m and the number of measurement $I=16384$.}}
  \label{image2}
\end{figure*}

\begin{figure}[t]
  \centering  \centerline{\includegraphics[width=7.4cm,height=5.5cm]{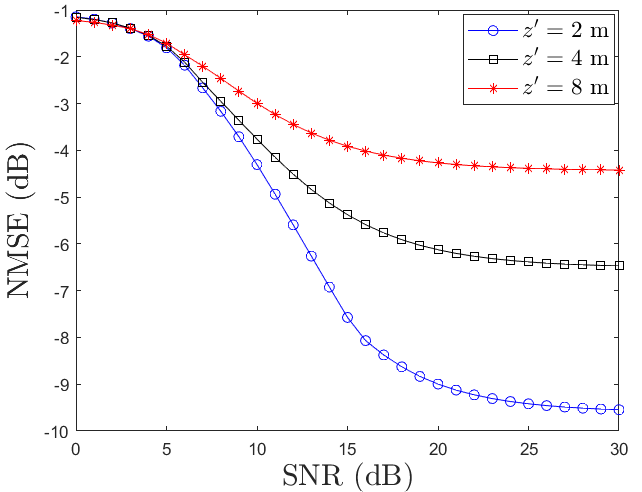}}  \caption{\textcolor{black}{Imaging performance versus SNR with $I=4096$}}
  \label{nmse2}
\end{figure}

To illustrate the imaging result vividly, we present the reconstructed images of two target patterns when $z'=4$~m and $I=16384$.
As shown in Fig.~\ref{image2}, when SNR $=10$~dB, the high-spatial-frequency speckles caused by noise cover the reconstructed image, making the target almost unrecognizable.
When SNR $=30$~dB, the reconstructed images are generally in good agreements with the targets.
The only difference is the slight discrepancy between the background of the image 
and that of the original target.

The NMSE versus SNR with different $z'$ is shown in Fig.~\ref{nmse2}, where $I=4096$ is selected,  \textcolor{black}{and the target ``SEU" is used in test}.
It is seen that, the NMSE decreases with the increase of SNR for all $z'$, and the best performance is achieved when $z'=2$~m.
When SNR is smaller than $5$~dB, the NMSE is large and decreases slowly with the increase of SNR. 
In this stage, the differences among the NMSE for different $z'$ are not pronounced. 
The NMSE decreases rapidly when SNR increases from $5$~dB to $20$~dB.
When SNR reaches the threshold of $20$~dB, the NMSE decreases to the error floor and no longer changes. 
At this point, the restriction of imaging quality is no longer the noise at the receiver, but is the lack of the orthogonality among the masks used for each measurement.
As the target gets farther away from the RIS, the discrepancy between the practical EM field distribution patterns and the ideal Hadamard masks becomes greater.
Thus, the orthogonality among the actual generated virtual masks becomes weaker,
leading to a larger error floor in imaging.
Moreover, an error floor is reached at lower SNR levels.

\subsection{Imaging Performance for a 3D Inhomogeneous Target}
\begin{figure*}[t]
  \centering
\begin{minipage}[t]{0.32\linewidth}
\subfigure[real relative permittivity]{
\includegraphics[width=5.5cm]{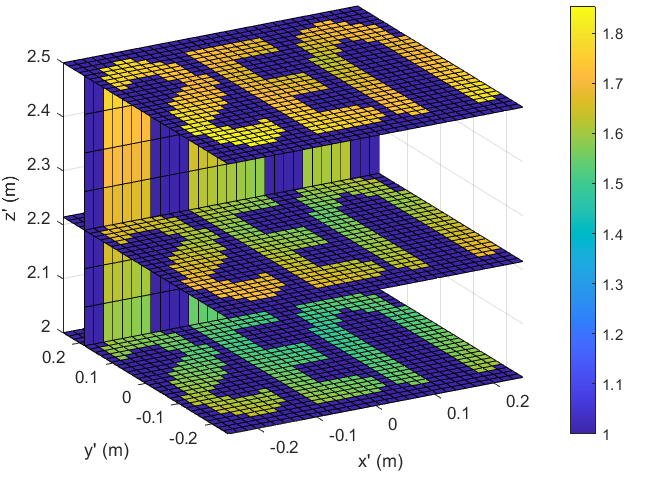}} 
\end{minipage}
\begin{minipage}[t]{0.32\linewidth}
\subfigure[recovered relative permittivity with SNR = 10~dB]{
\includegraphics[width=5.5cm]{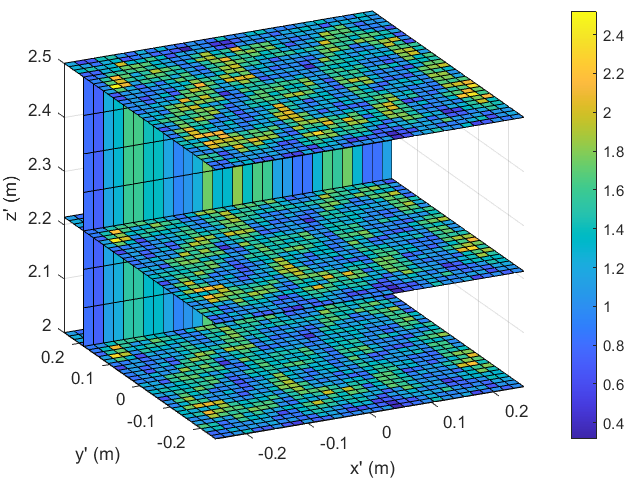}} 
\end{minipage}
\begin{minipage}[t]{0.32\linewidth}
\subfigure[recovered relative permittivity with SNR = 30~dB]{
\includegraphics[width=5.5cm]{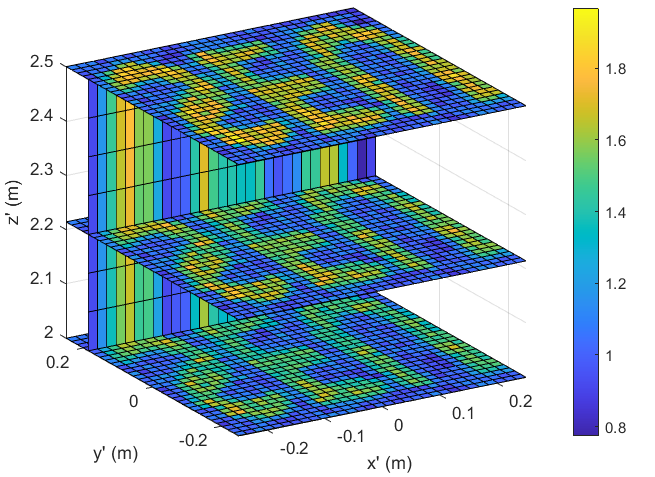}} 
\end{minipage} \\
\begin{minipage}[t]{0.32\linewidth}
\subfigure[real conductivity]{
\includegraphics[width=5.5cm]{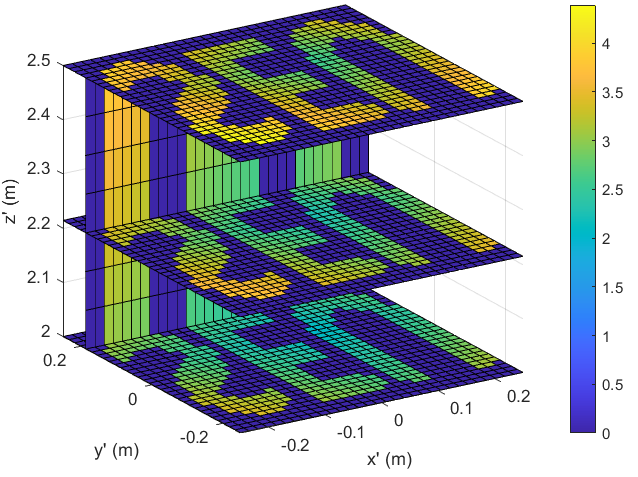}}
\end{minipage}
\begin{minipage}[t]{0.32\linewidth}
\subfigure[recovered conductivity with SNR = 10~dB]{
\includegraphics[width=5.5cm]{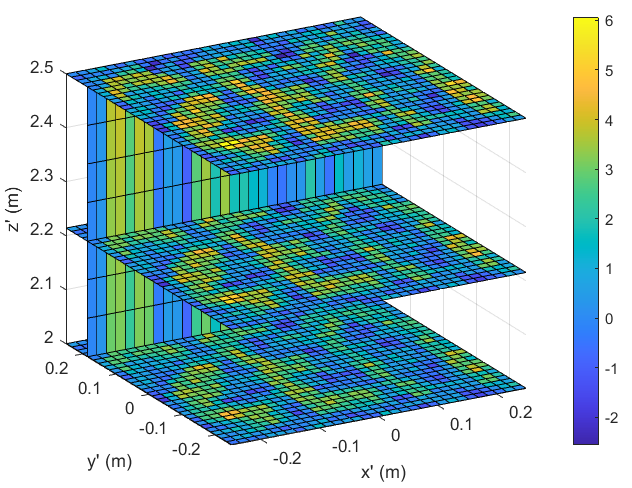}}
\end{minipage}
\begin{minipage}[t]{0.32\linewidth}
\subfigure[recovered conductivity with SNR = 30~dB]{
\includegraphics[width=5.5cm]{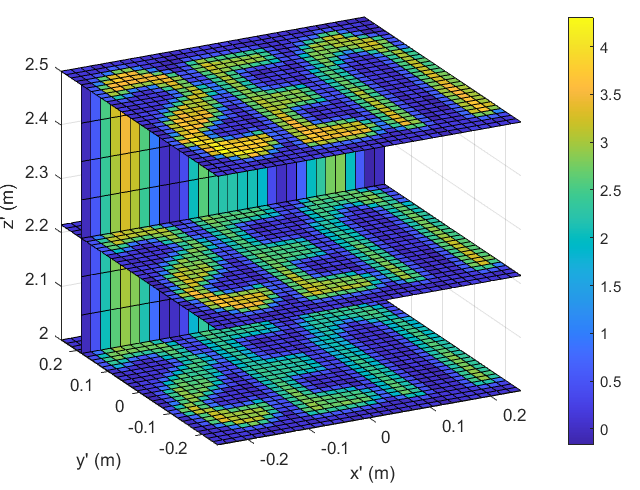}} 
\end{minipage}
\caption{\textcolor{black}{3D imaging performance with $z' \in [2,2.5] $~m and $I = 32768$, where the unit of conductivity is mS/m.
}}
\label{image3d}
\end{figure*}

The imaging domain $D$ is chosen as a cubic box whose length is $0.5$~m in each dimension.
We choose the number of sampled points on RIS as $N = 128 \times 128 = 16384$ and the number of sampled points in $D$ as $M = 32 \times 32 \times 8 = 8192$, where there are $8$ discrete slices along the $z$ axis. 
The Tikhonov regularization parameter is set as $\gamma= 10^{-14}$.
In order to compare with the images of the 2D scenario, the 3D target is columnar ``SEU" made of inhomogeneous materials as shown in Fig.~\ref{image3d}(a) and Fig.~\ref{image3d}(d). 
Both the relative permittivity and conductivity of the target increase with the increase of $z'$ and change linearly 
along the $x$ and $y$ directions. 

\begin{figure}[t]
  \centering  \centerline{\includegraphics[width=7.4cm,height=5.5cm]{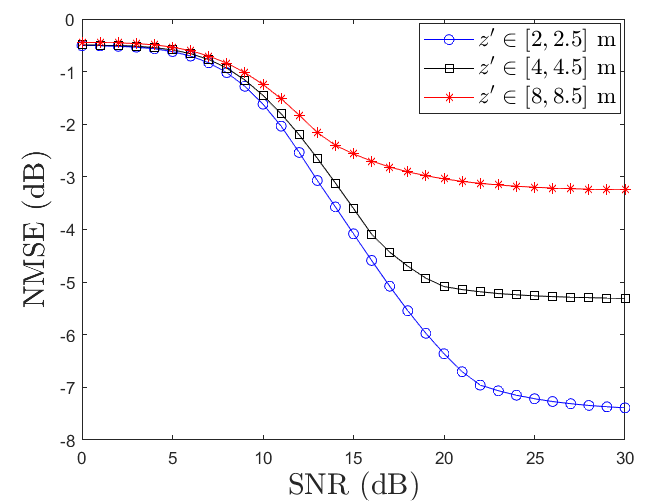}}
  \caption{\textcolor{black}{Imaging performance versus SNR with $I = 32768$}}
  \label{nmse3d}
\end{figure}


\textcolor{black}{
We present the reconstructed images through the relative permittivity and conductivity on three cross section planes
with $z' \in [2,2.5]$~m and $I=32768$.  
As shown in Fig.~\ref{image3d}, the reconstructed images reflect the approximate shape of the target on each cross section plane. 
When SNR $=10$~dB, the edges of the images are blurry and the reconstructed values of relative permittivity and conductivity have a large deviation.
There are points not in line with physical reality, i.e., with relative permittivity less than 1 or negative conductivity. 
Besides, the differences among the images on the three planes are not clearly reflected.
When SNR $=30$~dB, the edges are sharper and the reconstructed images are generally in good agreements with the targets. 
Moreover, the changes of relative permittivity and conductivity along $z$ axis are resolved in the reconstructed images. 
In general, compared to the images of the 2D homogeneous target, the images are more blurry and the contrast of images is less significant due to the inaccuracy of BA. }

\textcolor{black}{
The NMSE versus SNR is shown in Fig.~\ref{nmse3d} with $I = 32768$. 
It is seen that, the NMSE decreases with the increase of SNR, and the best performance is achieved when $z' \in [2,2.5]$~m. 
When SNR is smaller than $10$~dB, the NMSE is large and decreases slowly with the increase of SNR. 
In this stage, the differences among the NMSE for different $z'$ are not pronounced. 
When SNR reaches a threshold, the NMSE decreases to an error floor.   
The error floor mainly comes from the mismatch between BA and the real scattering process as well as the lack of orthogonality among the actual generated virtual masks.  
Since the 3D image is composed of more sampled points compared to the 2D image, the number of the measurements is larger correspondingly. 
However, the NMSE of the 3D target is still generally larger than that of the 2D homogeneous target due to the disparity
between BA and the real scattering process of the 3D inhomogeneous target.
}





\section{Conclusion}
In this paper, we propose a novel imaging scheme that employs holographic active RIS to create virtual EM masks on the target. 
This method simplifies the process and reduces the computation costs compared to the conventional methods.
To implement the approach, the RIS generates various virtual EM masks on the target, and the cross-correlation between the mask patterns and the electric field strength at the receiver is then calculated to extract information from the target and produce high-quality images.
We then propose a RIS design criterion for generating the proposed virtual EM masks.
\textcolor{black}{
The simulation results demonstrate that the proposed method is effective and robust for both 2D and 3D targets.
} 
Moreover, the quality of the images could be further enhanced by properly adjusting SNR, the number of generated masks, as well as the distance between the target and the RIS.



 \small 
 \bibliographystyle{ieeetr}
 \bibliography{IEEEabrv,mainbib}

\begin{thebibliography}{10}

\bibitem{Sheen:3DmmWaveImaging}
D.~M. Sheen, D.~L. McMakin, and T.~E. Hall, ``Three-dimensional millimeter-wave
  imaging for concealed weapon detection,'' {\em IEEE Trans. Microw. Theory
  Tech.}, vol.~49, pp.~1581--1592, Sep. 2001.

\bibitem{9827908}
D.~Chen, S.~Vakalis, and J.~A. Nanzer, ``Millimeter-wave imaging using a
  rotating dynamic antenna array and noise illumination,'' {\em IEEE Trans.
  Antennas Propag.}, vol.~70, no.~11, pp.~10965--10973, 2022.

\bibitem{8668512}
J.~M. Felício, J.~M. Bioucas-Dias, J.~R. Costa, and C.~A. Fernandes, ``Antenna
  design and near-field characterization for medical microwave imaging
  applications,'' {\em IEEE Trans. Antennas Propag.}, vol.~67, no.~7,
  pp.~4811--4824, 2019.

\bibitem{9743695}
O.~Fiser, V.~Hruby, J.~Vrba, T.~Drizdal, J.~Tesarik, J.~Vrba~Jr, and D.~Vrba,
  ``{UWB} bowtie antenna for medical microwave imaging applications,'' {\em
  IEEE Trans. Antennas Propag.}, vol.~70, no.~7, pp.~5357--5372, 2022.

\bibitem{4148072}
H.~M. Jafari, M.~J. Deen, S.~Hranilovic, and N.~K. Nikolova, ``A study of
  ultrawideband antennas for near-field imaging,'' {\em IEEE Trans. Antennas
  Propag.}, vol.~55, no.~4, pp.~1184--1188, 2007.

\bibitem{rubaek2007nonlinear}
T.~Rub{\ae}k, P.~M. Meaney, P.~Meincke, and K.~D. Paulsen, ``Nonlinear
  microwave imaging for breast-cancer screening using gauss--newton's method
  and the cgls inversion algorithm,'' {\em IEEE Trans. Antennas Propag.},
  vol.~55, no.~8, pp.~2320--2331, 2007.

\bibitem{li2005overview}
X.~Li, E.~J. Bond, B.~D. Van~Veen, and S.~C. Hagness, ``An overview of
  ultra-wideband microwave imaging via space-time beamforming for early-stage
  breast-cancer detection,'' {\em IEEE Antennas and Propagation Magazine},
  vol.~47, no.~1, pp.~19--34, 2005.

\bibitem{sadeghi2019dort}
S.~Sadeghi, K.~Mohammadpour-Aghdam, R.~Faraji-Dana, and R.~J. Burkholder, ``A
  dort-uniform diffraction tomography algorithm for through-the-wall imaging,''
  {\em IEEE Trans. Antennas Propag.}, vol.~68, no.~4, pp.~3176--3183, 2019.

\bibitem{ali20103d}
M.~A. Ali and M.~Moghaddam, ``3{D} nonlinear super-resolution microwave
  inversion technique using time-domain data,'' {\em IEEE Trans. Antennas
  Propag.}, vol.~58, no.~7, pp.~2327--2336, 2010.

\bibitem{ren2016uniform}
K.~Ren and R.~J. Burkholder, ``A uniform diffraction tomographic imaging
  algorithm for near-field microwave scanning through stratified media,'' {\em
  IEEE Trans. Antennas Propag.}, vol.~64, no.~12, pp.~5198--5207, 2016.

\bibitem{scan}
R.~K. Amineh, M.~Ravan, A.~Khalatpour, and N.~K. Nikolova, ``Three-dimensional
  near-field microwave holography using reflected and transmitted signals,''
  {\em IEEE Trans. Antennas Propag.}, vol.~59, no.~12, pp.~4777--4789, 2011.

\bibitem{molaei2022development}
A.~M. Molaei, T.~Fromenteze, V.~Skouroliakou, T.~V. Hoang, R.~Kumar, V.~Fusco,
  and O.~Yurduseven, ``Development of fast fourier-compatible image
  reconstruction for 3{D} near-field bistatic microwave imaging with dynamic
  metasurface antennas,'' {\em IEEE Trans. Vehi. Tech.}, vol.~71, no.~12,
  pp.~13077--13090, 2022.

\bibitem{RIS_image}
Y.~Tao and Z.~Zhang, ``Distributed computational imaging with reconfigurable
  intelligent surface,'' in {\em 2020 International Conference on Wireless
  Communications and Signal Processing (WCSP)}, pp.~448--454, Oct. 2020.

\bibitem{9384499}
J.~Liu, J.~Zhang, Q.~Zhang, J.~Wang, and X.~Sun, ``Secrecy rate analysis for
  reconfigurable intelligent surface-assisted {MIMO} communications with
  statistical {CSI},'' {\em China Commun.}, vol.~18, no.~3, pp.~52--62, Mar.
  2021.

\bibitem{9149203}
F.~Yang, J.-B. Wang, H.~Zhang, C.~Chang, and J.~Cheng, ``Intelligent reflecting
  surface-assisted mmwave communication exploiting statistical csi,'' in {\em
  Proc. IEEE ICC}, pp.~1--6, Mar. 2020.

\bibitem{8746155}
Y.~Han, W.~Tang, S.~Jin, C.-K. Wen, and X.~Ma, ``Large intelligent
  surface-assisted wireless communication exploiting statistical csi,'' {\em
  IEEE Trans. Vehi. Tech.}, vol.~68, no.~8, pp.~8238--8242, Aug. 2019.

\bibitem{9145334}
J.~Zhang, J.~Liu, S.~Ma, C.-K. Wen, and S.~Jin, ``Transmitter design for large
  intelligent surface-assisted {MIMO} wireless communication with statistical
  {CSI},'' in {\em Proc. IEEE ICC Workshops}, pp.~1--5, Mar. 2020.

\bibitem{5}
O.~Ozdogan, E.~Bjornson, and E.~G. Larsson, ``Intelligent reflecting surfaces:
  Physics, propagation, and pathloss modeling,'' {\em IEEE Wireless Commun.
  Lett.}, vol.~9, no.~5, pp.~581--585, May. 2020.

\bibitem{6}
M.~Najafi, V.~Jamali, R.~Schober, and H.~V. Poor, ``Physics-based modeling and
  scalable optimization of large intelligent reflecting surfaces,'' {\em IEEE
  Trans. Wireless Commun.}, vol.~69, no.~4, pp.~2673--2691, Apr. 2021.

\bibitem{7}
N.~Mohammadi~Estakhri and A.~Al\`u, ``Wave-front transformation with gradient
  metasurfaces,'' {\em Phys. Rev. X}, vol.~6, p.~041008, Oct 2016.

\bibitem{8}
F.~H. Danufane, M.~D. Renzo, J.~de~Rosny, and S.~Tretyakov, ``On the path-loss
  of reconfigurable intelligent surfaces: An approach based on green’s
  theorem applied to vector fields,'' {\em IEEE Trans. Wireless Commun.},
  vol.~69, no.~8, pp.~5573--5592, Aug. 2021.

\bibitem{RIS}
C.~Huang, A.~Zappone, G.~C. Alexandropoulos, M.~Debbah, and C.~Yuen,
  ``Reconfigurable intelligent surfaces for energy efficiency in wireless
  communication,'' {\em IEEE Trans. Wireless Commun.}, vol.~18, no.~8,
  pp.~4157--4170, 2019.

\bibitem{9343768}
N.~S. Perović, L.-N. Tran, M.~Di~Renzo, and M.~F. Flanagan, ``Achievable rate
  optimization for {MIMO} systems with reconfigurable intelligent surfaces,''
  {\em IEEE Trans. Wireless Commun.}, vol.~20, no.~6, pp.~3865--3882, Jan.
  2021.

\bibitem{near_imaging}
J.~Han, L.~Li, S.~Tian, X.~Ma, Q.~Feng, H.~Liu, Y.~Zhao, and G.~Liao,
  ``Frequency-diverse holographic metasurface antenna for near-field microwave
  computational imaging,'' {\em Frontiers in Materials}, vol.~8, Oct. 2021.

\bibitem{9}
O.~Rinchi, A.~Elzanaty, and M.-S. Alouini, ``Compressive near-field
  localization for multipath ris-aided environments,'' {\em IEEE Commun.
  Lett.}, vol.~26, no.~6, pp.~1268--1272, Aug. 2022.

\bibitem{10}
X.~Wei, L.~Dai, Y.~Zhao, G.~Yu, and X.~Duan, ``Codebook design and beam
  training for extremely large-scale ris: Far-field or near-field?,'' {\em
  China Commun.}, vol.~19, no.~6, pp.~193--204, Feb. 2022.

\bibitem{mypaper}
Y.~Jiang, F.~Gao, M.~Jian, S.~Zhang, and W.~Zhang, ``Reconfigurable intelligent
  surface for near field communications: Beamforming and sensing,'' {\em IEEE
  Trans. Wireless Commun.}, pp.~1--1, Nov. 2022.

\bibitem{h1}
Z.~Wan, Z.~Gao, F.~Gao, M.~Di~Renzo, and M.-S. Alouini, ``Terahertz massive
  mimo with holographic reconfigurable intelligent surfaces,'' {\em IEEE Trans.
  Commun.}, vol.~69, no.~7, pp.~4732--4750, July 2021.

\bibitem{h2}
C.~Huang, S.~Hu, G.~C. Alexandropoulos, A.~Zappone, C.~Yuen, R.~Zhang,
  M.~Di~Renzo, and M.~Debbah, ``Holographic mimo surfaces for 6g wireless
  networks: Opportunities, challenges, and trends,'' {\em IEEE Wireless
  Commun.}, vol.~27, no.~5, pp.~118--125, July 2020.

\bibitem{antenna}
G.~Wu, K.~F. Chan, K.~M. Shum, and C.~H. Chan, ``Millimeter-wave holographic
  flat lens antenna for orbital angular momentum multiplexing,'' {\em IEEE
  Trans. Antennas Propag.}, vol.~69, pp.~4289--4303, 2021.

\bibitem{h3}
A.~Pizzo, T.~L. Marzetta, and L.~Sanguinetti, ``Spatially-stationary model for
  holographic mimo small-scale fading,'' {\em IEEE J. Sele. Areas Commun.},
  vol.~38, no.~9, pp.~1964--1979, Jan. 2020.

\bibitem{h4}
R.~Deng, B.~Di, H.~Zhang, Y.~Tan, and L.~Song, ``Reconfigurable holographic
  surface: Holographic beamforming for metasurface-aided wireless
  communications,'' {\em IEEE Trans. Vehi. Tech.}, vol.~70, no.~6,
  pp.~6255--6259, Jan. 2021.

\bibitem{inverse}
M.~Bertero, P.~Boccacci, and C.~De~Mol, {\em Introduction to inverse problems
  in imaging}.
\newblock CRC press, 2021.

\bibitem{9424177}
Y.~Liu, X.~Liu, X.~Mu, T.~Hou, J.~Xu, M.~Di~Renzo, and N.~Al-Dhahir,
  ``Reconfigurable intelligent surfaces: Principles and opportunities,'' {\em
  IEEE Commun. Surv. Tutorials}, vol.~23, no.~3, pp.~1546--1577, Oct. 2021.

\bibitem{9110889}
B.~Di, H.~Zhang, L.~Song, Y.~Li, Z.~Han, and H.~V. Poor, ``Hybrid beamforming
  for reconfigurable intelligent surface based multi-user communications:
  Achievable rates with limited discrete phase shifts,'' {\em IEEE J. Sele.
  Areas Commun.}, vol.~38, no.~8, pp.~1809--1822, Aug. 2020.

\bibitem{active1}
K.~Zhi, C.~Pan, H.~Ren, K.~K. Chai, and M.~Elkashlan, ``Active ris versus
  passive ris: Which is superior with the same power budget?,'' {\em IEEE
  Commun. Lett.}, vol.~26, no.~5, pp.~1150--1154, Mar. 2022.

\bibitem{active2}
Z.~Zhang, L.~Dai, X.~Chen, C.~Liu, F.~Yang, R.~Schober, and H.~V. Poor,
  ``Active ris vs. passive ris: Which will prevail in 6g?,'' {\em IEEE Trans.
  Commun.}, Dec. 2022.

\bibitem{active3}
H.~Niu, Z.~Lin, K.~An, X.~Liang, Y.~Hu, D.~Li, and G.~Zheng, ``Active
  ris-assisted secure transmission for cognitive satellite terrestrial
  networks,'' {\em IEEE Trans. Vehi. Tech.}, Sep. 2022.

\bibitem{active4}
Z.~Peng, X.~Liu, C.~Pan, L.~Li, and J.~Wang, ``Multi-pair {D}2{D}
  communications aided by an active ris over spatially correlated channels with
  phase noise,'' {\em IEEE Wireless Commun. Lett.}, vol.~11, no.~10,
  pp.~2090--2094, July 2022.

\bibitem{2d}
M.~Ravan, R.~K. Amineh, and N.~K. Nikolova, ``Two-dimensional near-field
  microwave holography,'' {\em Inverse Problems}, vol.~26, p.~055011, Apr.
  2010.

\bibitem{soft}
C.~Liaskos, S.~Nie, A.~Tsioliaridou, A.~Pitsillides, S.~Ioannidis, and
  I.~Akyildiz, ``A new wireless communication paradigm through
  software-controlled metasurfaces,'' {\em IEEE Commun. Mag.}, vol.~56, no.~9,
  pp.~162--169, Sep. 2018.

\bibitem{textbook}
C.~A. Balanis, {\em Advanced Engineering Electromagnetics}.
\newblock USA: Wiley: Hoboken, NJ, 2~ed., 2012.

\bibitem{lipp}
J.~O. Vargas and R.~Adriano, ``Subspace-based conjugate-gradient method for
  solving inverse scattering problems,'' {\em IEEE Trans. Antennas Propag.},
  vol.~70, no.~12, pp.~12139--12146, 2022.

\bibitem{lipp2}
F.-F. Wang and Q.~H. Liu, ``A hybrid {B}orn iterative {B}ayesian inversion
  method for electromagnetic imaging of moderate-contrast scatterers with
  piecewise homogeneities,'' {\em IEEE Trans. Antennas Propag.}, vol.~70,
  no.~10, pp.~9652--9661, 2022.

\bibitem{born}
D.~Tajik, R.~Kazemivala, and N.~K. Nikolova, ``Real-time imaging with
  simultaneous use of born and rytov approximations in quantitative microwave
  holography,'' {\em IEEE Trans. Microw. Theory Tech.}, vol.~70, no.~3,
  pp.~1896--1909, 2022.

\bibitem{cor}
J.~H. Shapiro, ``Computational ghost imaging,'' {\em Phys. Rev. A}, vol.~78,
  p.~061802, Dec 2008.

\bibitem{linear_algebra}
J.~W. Demmel, {\em Applied Numerical Linear Algebra}.
\newblock Society for Industrial and Applied Mathematics, 1997.

\bibitem{cor2}
Y.~Bromberg, O.~Katz, and Y.~Silberberg, ``Ghost imaging with a single
  detector,'' {\em Phys. Rev. A}, vol.~79, p.~053840, May 2009.

\bibitem{cor3}
M.~Braasch, V.~F. Gili, T.~Pertsch, and F.~Setzpfandt, ``Classical ghost
  imaging: A comparative study of algorithmic performances for image
  reconstruction in prospect of plenoptic imaging,'' {\em IEEE Photonics
  Journal}, vol.~13, no.~3, pp.~1--14, 2021.

\bibitem{cor4}
J.~H. Shapiro and R.~W. Boyd, ``The physics of ghost imaging,'' {\em Quantum
  Information Processing}, vol.~11, no.~4, pp.~949--993, 2012.

\bibitem{cor5}
B.~I. Erkmen and J.~H. Shapiro, ``Ghost imaging: from quantum to classical to
  computational,'' {\em Advances in Optics and Photonics}, vol.~2, no.~4,
  pp.~405--450, 2010.

\bibitem{orthogonal}
P.~Cohen, C.~LeDinh, and V.~Lacasse, ``Classification of natural textures by
  means of two-dimensional orthogonal masks,'' {\em IEEE Transactions on
  Acoustics, Speech, and Signal Processing}, vol.~37, no.~1, pp.~125--128, Apr.
  1989.

\bibitem{infeasible}
R.~Deng, B.~Di, H.~Zhang, D.~Niyato, Z.~Han, H.~V. Poor, and L.~Song,
  ``Reconfigurable holographic surfaces for future wireless communications,''
  {\em IEEE Wireless Commun.}, vol.~28, no.~6, pp.~126--131, 2021.

\end{thebibliography}

 \begin{IEEEbiography}[{\includegraphics[width=1in,height=1.25in,clip,keepaspectratio]{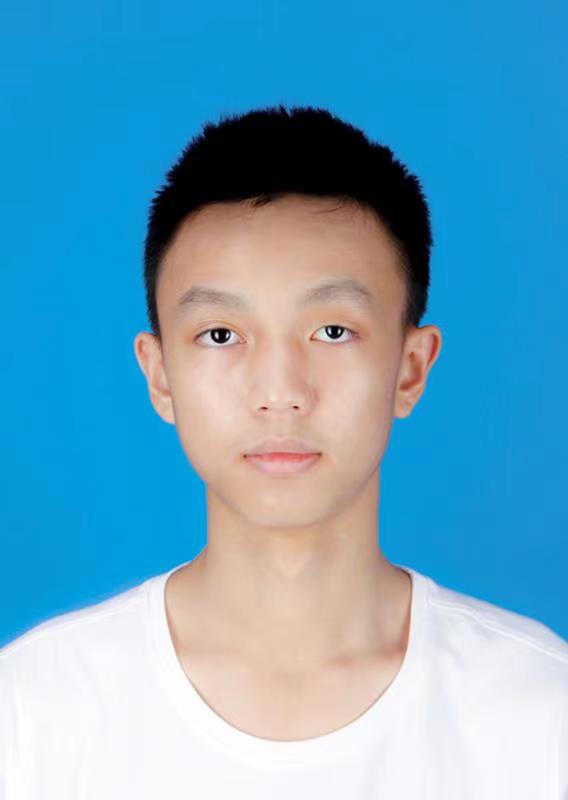}}]
 {Yuhua Jiang} is now an undergraduate student in Tsinghua University. His research interests include 
 integrated sensing and communication, electromagnetic property sensing, massive MIMO, reconfigurable intelligent surface, and electromagnetic information theory. 
\end{IEEEbiography}

\begin{IEEEbiography}[{\includegraphics[width=1in,height=1.25in,clip,keepaspectratio]{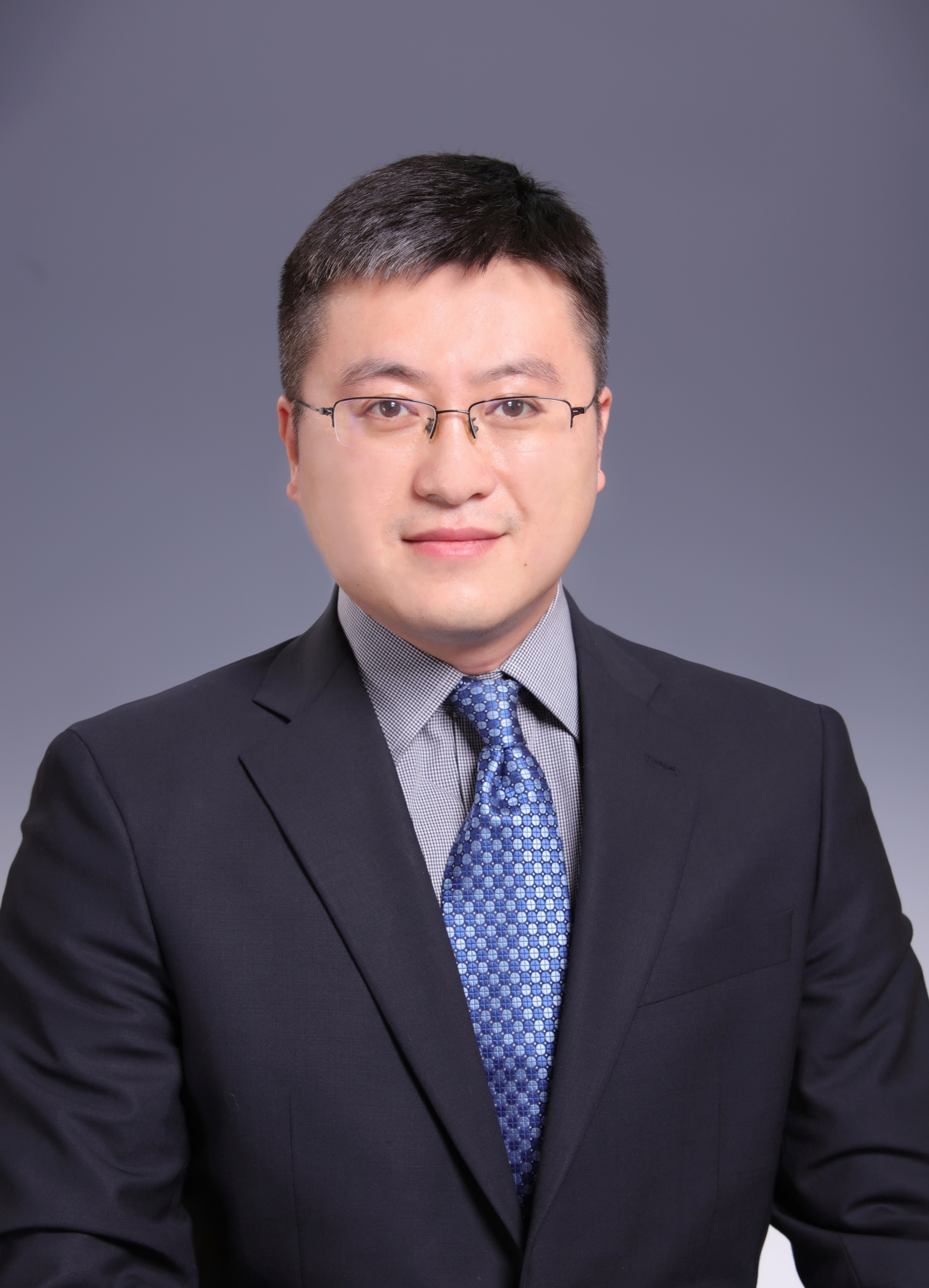}}]
 {Feifei Gao} (Fellow, IEEE) received the B.Eng. degree from Xi'an Jiaotong University, Xi'an, China in 2002, the M.Sc. degree from McMaster University, Hamilton, ON, Canada in 2004, and the Ph.D. degree from National University of Singapore, Singapore in 2007. Since 2011, he joined the Department of Automation, Tsinghua University, Beijing, China, where he is currently a Professor. 

Prof. Gao's research interests include signal processing for communications, array signal processing, convex optimizations, and artificial intelligence assisted communications. He has authored/coauthored more than 200 refereed IEEE journal papers and more than 150 IEEE conference proceeding papers that are cited more than 17000 times in Google Scholar. Prof. Gao has served as an Editor of IEEE Transactions on Wireless Communications, IEEE Journal of Selected Topics in Signal Processing (Lead Guest Editor), IEEE Transactions on Cognitive Communications and Networking, IEEE Signal Processing Letters (Senior Editor), IEEE Communications Letters (Senior Editor), IEEE Wireless Communications Letters, and China Communications. He has also served as the symposium co-chair for 2019 IEEE Conference on Communications (ICC), 2018 IEEE Vehicular Technology Conference Spring (VTC), 2015 IEEE Conference on Communications (ICC), 2014 IEEE Global Communications Conference (GLOBECOM), 2014 IEEE Vehicular Technology Conference Fall (VTC), as well as Technical Committee Members for more than 50 IEEE conferences.
\end{IEEEbiography}

 \begin{IEEEbiography}[{\includegraphics[width=1in,height=1.25in,clip,keepaspectratio]{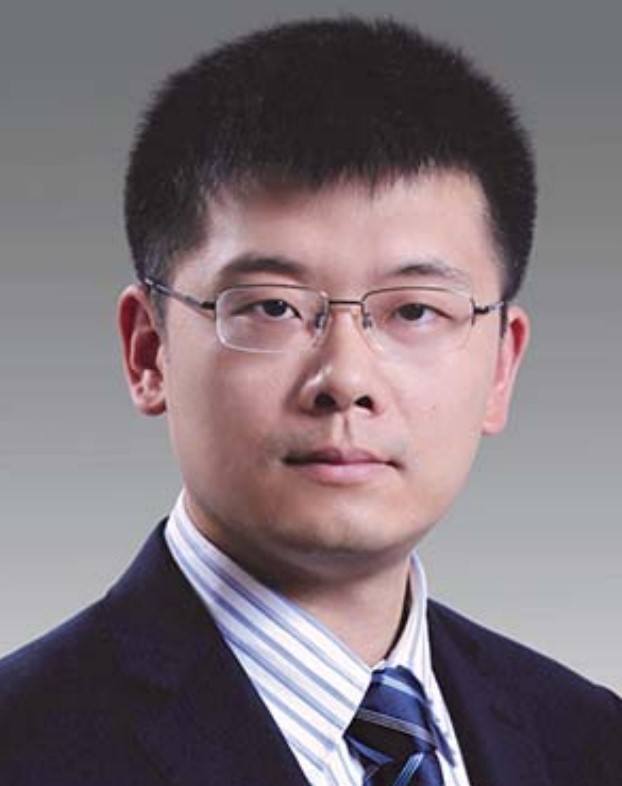}}]
 {Yimin Liu} (Member, IEEE) received the B.S. and Ph.D. degrees (Hons.) in electronics engineering from Tsinghua University, China, in 2004 and 2009, respectively, where he was with the Intelligence Sensing Lab, Department of Electronic Engineering in 2004. He is currently an Associate Professor with Tsinghua University, where his field of activity is study on new concept radar and other microwave sensing technologies. His current research interests include radar theory, statistic signal processing, compressive sensing and their applications in radar, spectrum sensing, and intelligent transportation systems. 
\end{IEEEbiography}

 \begin{IEEEbiography}[{\includegraphics[width=1in,height=1.25in,clip,keepaspectratio]{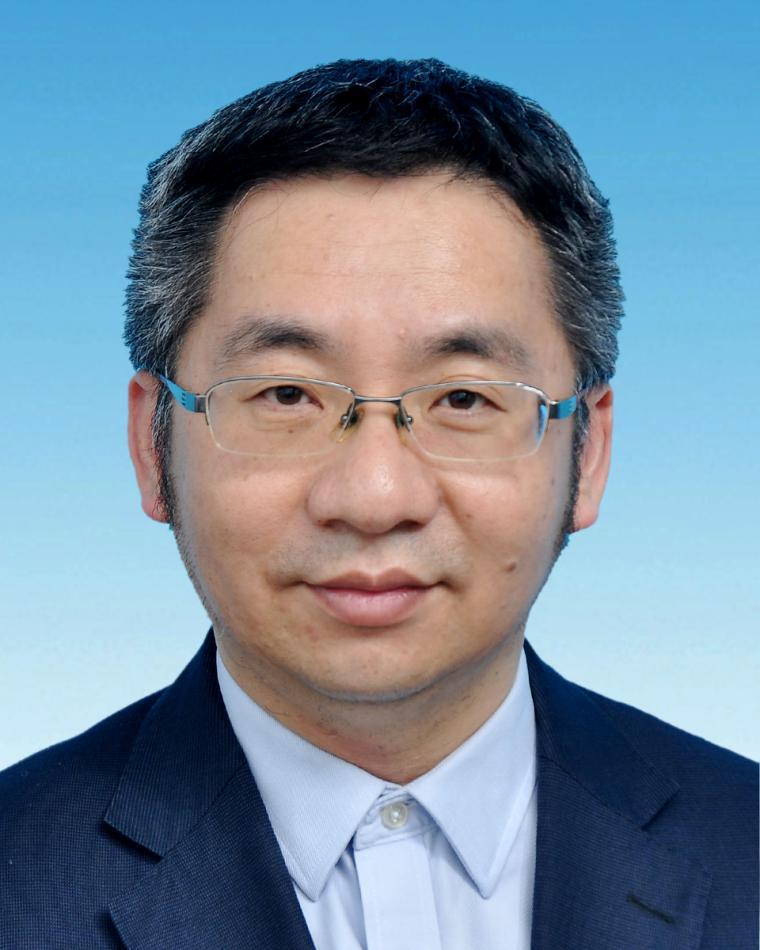}}]
 {Shi Jin} (Fellow, IEEE) received the B.S. degree in communications engineering from the Guilin University of Electronic Technology, Guilin, China, in 1996, the M.S. degree from the Nanjing University of Posts and Telecommunications, Nanjing, China, in 2003, and the Ph.D. degree in information and communications engineering from Southeast University, Nanjing, in 2007. From June 2007 to October 2009, he was a Research Fellow with the Adastral Park Research Campus, University College London, London, U.K. He is currently with the Faculty of the National Mobile Communications Research Laboratory, Southeast University. 

His research interests include wireless communications, random matrix theory, and information theory. He and his coauthors received the 2011 IEEE Communications Society Stephen O. Rice Prize Paper Award in the field of communication theory, the 2022 Best Paper Award, and the 2010 Young Author Best Paper Award by the IEEE Signal Processing Society. He is serving as an Area Editor for the e IEEE Transactions on Communications and IET Electronics Letters. He was an Associate Editor of the IEEE Transactions on Wireless Communications, IEEE Communications Letters, and IET Communications. 
\end{IEEEbiography}

 \begin{IEEEbiography}[{\includegraphics[width=1in,height=1.25in,clip,keepaspectratio]{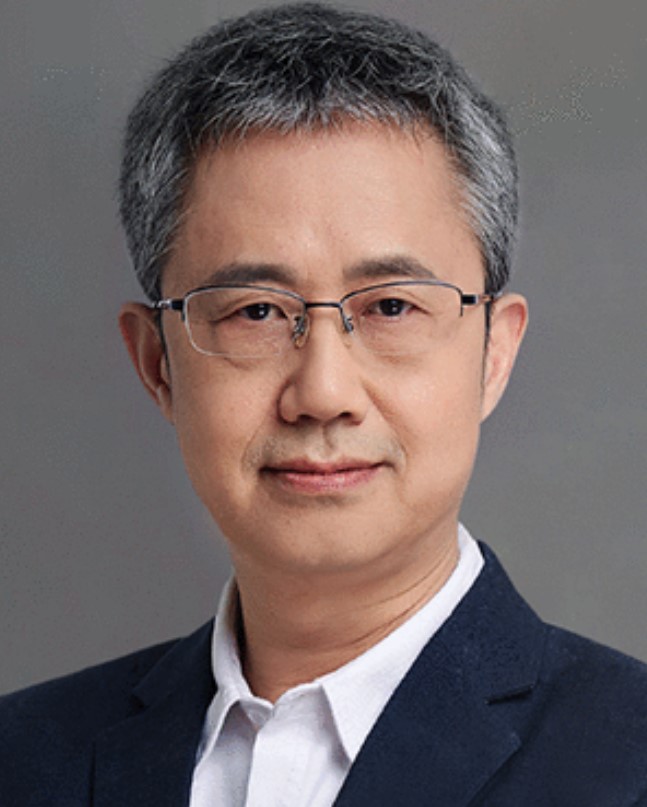}}]
 {Tiejun Cui} (Fellow, IEEE) received the B.Sc., M.Sc., and Ph.D. degrees in electrical engineering from Xidian University, Xi’an, China, in 1987, 1990, and 1993, respectively.

In March 1993, he joined the Department of Electromagnetic Engineering, Xidian University, and was promoted to an Associate Professor in November 1993. From 1995 to 1997, he was a Research Fellow with the Institut fur Hochstfrequenztechnik und Elektronik (IHE), University of Karlsruhe, Karlsruhe, Germany. In July 1997, he joined the Center for Computational Electromagnetics, Department of Electrical and Computer Engineering, University of Illinois at Urbana-Champaign, Urbana, IL, USA, first as a Post-Doctoral Research Associate and then a Research Scientist. In September 2001, he was a Cheung-Kong Professor with the Department of Radio Engineering, Southeast University, Nanjing, China. He is currently the Chief Professor of Southeast University, the Director of the State Key Laboratory of Millimeter Waves, and the Founding Director of the Institute of Electromagnetic Space, Southeast University. He proposed the concepts of digital coding and programmable metamaterials and realized their first prototypes, based on which he founded the new direction of information metamaterials, bridging the physical world and digital world. He is the first author of books Metamaterials: Theory, Design, and Applications (Springer, November 2009); Metamaterials: Beyond Crystals, Noncrystals, and Quasicrystals (CRC Press, March 2016); and Information Metamaterials (Cambridge University Press, 2021). He has published over 500 peer-review journal papers, which have been cited by more than 44 600 times (H-factor 107, Google Scholar), and licensed over 100 patents. His research interests include metamaterials and computational electromagnetics. 
\end{IEEEbiography}

\ifCLASSOPTIONcaptionsoff
  \newpage
\fi



%

\end{document}